\RequirePackage{ifpdf}
\documentclass[a4paper,11pt]{article}
\pdfoutput=1
\usepackage{jheppub}
\usepackage{amsmath,amsbsy,amssymb} 
\usepackage{graphicx}
\usepackage{epsfig}
\usepackage{fixmath}
\usepackage{caption}
\usepackage{subcaption}
\usepackage{slashed}
\usepackage{xspace}
\graphicspath{plots/} % TeX will look for pix here

\newcommand\new{\newcommand}         % shorthand for \newcommand
\newcommand{\mt      }{\ensuremath{m_{t}}\xspace}
\newcommand{\mtin    }{\ensuremath{m^{\mathrm{in}}_{t}}\xspace}
\newcommand{\mtou    }{\ensuremath{m^{\mathrm{out}}_{t}}\xspace}
\newcommand{\mlb     }{\ensuremath{m_{lb}}\xspace}
\def\bi{\begin{itemize}}   
\def\ei{\end{itemize}}
\def\be{\begin{equation}}   
\def\ee{\end{equation}}
\def\bea{\begin{eqnarray}}  
\def\eea{\end{eqnarray}}

\def\MSbar{$\overline{{\rm MS}}$}

\def\gev{\mathrm{\:GeV}}
\def\tev{\mathrm{\:TeV}}
\def\tsc{\textsc}
\def\trm{\textrm}
\def\mrm{\mathrm}
\def\srm#1{{\trm{\tiny #1}}}

\new{\as}[1]      {{\ifmmode\alpha^{#1}_s
                    \else$\alpha^{#1}_s$\fi}}
\new{\lqcd}       {{\ifmmode\Lambda_{\mathrm{ QCD}}
                    \else $\Lambda_{\mathrm{ QCD}}$\fi}}

\new{\mpi}{Max-Planck-Institut f\"ur Physik, %
  F\"ohringer Ring 6, %
  80805 M\"unchen, %
  Germany}

\title{NLO QCD corrections to $\mathbold{W^+W^-b\bar b}$ production
  with leptonic decays in the light of top quark mass and asymmetry
  measurements}

\author{Gudrun~Heinrich, Andreas~Maier, Richard~Nisius,
  Johannes~Schlenk and Jan~Winter\,$^a$} 
\affiliation[a]{\mpi}

\keywords{QCD, NLO Computations, LHC, Top Quark}

\abstract{
We present the NLO QCD corrections to the processes $pp$ and
$p\bar p\to W^+W^-b\bar b$ including leptonic decays of the $W$ bosons.
Non-resonant contributions as well as diagrams with doubly resonant
and singly resonant top quark propagators are fully taken into
account. We employ the narrow width approximation to perform the
decays of the $W$ bosons; spin correlations are however preserved. We
also calculate observables relevant for top quark mass measurements,
and study the impact of kinematical requirements and different scale
choices on $t\bar t$ asymmetries.}

\preprint{MPP-2013-318}

\begin{document}
%%%%%%%%%%%%%%%%%%%%%%%%%%%%%%%%%%%%%%%%%%%%%%%%%
%.newer.than.older.\clearpage
\maketitle
\section{Introduction}\label{sec:intro}
The production of top quarks is one of the most important reactions
studied at the Tevatron and the LHC. Especially the latter is known
for its status as a top quark factory, producing top quark pairs
copiously. Up to now more than $5\times10^6$ $t\bar t$ pairs have been
produced. Top quarks play a major role as a background to New Physics
searches, and also in precision studies which can provide indirect
hints to physics beyond the Standard Model.
Currently a lot of effort is put into the precise determination of the
top quark pair production cross section and differential distributions
such as the pair transverse momentum and the pair invariant mass.
Among the interesting observables are also the top quark mass
%~\cite{ATLAS-CONF-2013-077,ATLAS:2012aj,ATLAS-CONF-2013-046,Chatrchyan:2012ea,ATLAS-CONF-2013-102,Aaltonen:2011dr,Abazov:2012rp}
as well as the forward-backward and charge
asymmetry~\cite{Aaltonen:2011kc,Abazov:2011rq,CDF:2011vba,Aaltonen:2012it,CDF:2013gna,Aaltonen:2013vaf,Abazov:2013wxa,ATLAS:2012an,ATLAS-CONF-2012-057,Aad:2013cea,Chatrchyan:2011hk,Chatrchyan:2012cxa,CMS-PAS-TOP-12-010,CMS-PAS-TOP-12-033}
as measured at the Tevatron and LHC, respectively.

\smallskip
To match the experimental precision reached for quantities like the
top quark mass, the theory predictions need to go beyond the simple
approximation of factorizing top quark production and decay. For
example, finite width effects and non-factorizing contributions to
observables based on $W$ boson decay products and $b$ jets can have a
non-negligible impact on mass measurements. The latest combinations of
the top quark mass from the Tevatron and the LHC can be found in
Refs.~\cite{CDF:2013jga}~and~\cite{ATLAS-CONF-2012-095},
respectively. Also, the first combined measurement using ATLAS, CDF,
CMS and D\O~data has recently been published~\cite{ATLAS:2014wva}.
The contributing measurements relevant to the work of our paper are
those utilizing the leptonic decay mode. They are discussed
in~\cite{Abbott:1997fv,Abbott:1998dn,Abe:1998bf,Aaltonen:2009hd,Aaltonen:2011dr,Abazov:2012rp}
and~\cite{ATLAS-CONF-2013-077,Chatrchyan:2012ea} based on results
obtained at the Tevatron and the LHC, respectively.

\smallskip
The next-to-leading order (NLO) QCD corrections to top quark pair
production have been already known for a long
time~\cite{Nason:1987xz,Nason:1989zy,Beenakker:1990maa,Mangano:1991jk,%
Frixione:1995fj}. The NLO electroweak corrections were calculated
in \cite{Beenakker:1993yr}. Very recently the full NNLO cross section
for $t\bar{t}$ production has become available~\cite{Czakon:2013goa}.
These calculations treat the top quarks as stable on-shell particles. 
Decays can then be attached to the top quarks in the narrow width
approximation (NWA), where production and decay decouple. In most
applications, these decays are calculated only at the leading order.
One however makes use of spin density matrix or reweighting techniques
to preserve the spin correlations between particle production and
decay. This, especially, is the standard in multi-purpose Monte Carlo
event generators. At parton level, NLO calculations using the NWA were
further improved by promoting the treatment of top quark decays to
NLO.\footnote{One recent development presented in
  \cite{Bernreuther:2014dla} concerns the calculation of NLO
  corrections to polarized top quark decays with an additional jet in
  the final state.}
The complete evaluation of the $\mathcal{O}(\alpha_\mrm{s})$
corrections to $t\bar t$ production {\em and}\/ decay based on the NWA
and in full regard of spin correlations is documented in
Refs.~\cite{Melnikov:2009dn,Biswas:2010sa,Melnikov:2011ai}.

\smallskip
The full process $p\bar p$ or $pp\rightarrow W^+W^-b\bar{b}$ at
$\mathcal{O}(\alpha^2_\mrm{s}\alpha^2)$, where top quarks are treated
as off-shell particles, represents a $2\to4$ process which is
of much higher complexity. It includes resonant top quark production
and decay, but also singly resonant and non-resonant contributions.
Using massless $b$ quarks, this process was calculated at NLO in QCD
in~\cite{Denner:2010jp,Denner:2012yc,Bevilacqua:2010qb}. More
recently, as shown in ~\cite{Frederix:2013gra,Cascioli:2013wga}, it
was also computed in the 4-flavour scheme, i.e.~for massive $b$
quarks.

\smallskip
In this paper we calculate the NLO QCD corrections to the
$\mathcal{O}(\alpha^2_\mrm{s}\alpha^2)$ processes $p\bar p$ and
$pp\rightarrow W^+W^-b\bar{b}\rightarrow
(e^+ \nu_e)\,(\mu^- \bar{\nu}_{\mu})\,b\bar{b}$ in the 5-flavour scheme,
including singly resonant and non-resonant contributions,
corresponding to Feynman diagrams containing only one or no top quark
propagator that can go on-shell. The impact of non-resonant $W$
boson contributions has been studied in \cite{Denner:2012yc} and found
to be small. Therefore, non-resonant contributions from $W$ bosons are
neglected in our calculation.
On the other hand, in contrast to the calculations 
in \cite{Denner:2010jp,Denner:2012yc,Bevilacqua:2010qb}, 
contributions from (massless) $b$ quarks in the initial state are 
included in the calculation presented here.

\smallskip
The structure of this paper is as follows: in Section~\ref{sec:calc},
we give details about the calculation and present some numerical
results for LHC collisions at $7\tev$, in particular for observables
which are sensitive to the non-factorizing contributions.
In Section~\ref{sec:pheno}, we perform a detailed phenomenological
analysis of observables that are of particular interest for precision
studies: the top quark mass and observables related to $t\bar{t}$
asymmetries. Finally, we conclude in Section~\ref{sec:conclusion}.

%%%%%%%%%%%%%%%%%%%%%%%%%%%%%%%%%%%%%%%%%%%%%%%%%%

\section{Calculational framework and numerical results}\label{sec:calc}
For all our perturbative QCD, parton level calculations, we use the
\tsc{GoSam}~\cite{Cullen:2011ac} plus \tsc{Sherpa}~\cite{Gleisberg:2008ta}
combined generator package, in short \tsc{GoSam}+\tsc{Sherpa}. For
examples of applications, see
Refs.~\cite{vanDeurzen:2013rv,Hoeche:2013mua,Cullen:2013saa,vanDeurzen:2013xla},
for a list of pre-generated process packages, see~\cite{hepforge:proc}.
The multi-purpose Monte Carlo event generator \tsc{Sherpa} is used to
provide the Born, real radiation and subtraction term contributions,
as well as to accomplish the phase-space integration~\cite{Gleisberg:2008ta}.
The tree-level amplitudes are obtained from
both \tsc{Amegic}~\cite{Krauss:2001iv} and
\tsc{Comix}~\cite{Gleisberg:2008fv}, which are the \tsc{Sherpa} in-house
matrix-element generators, while the dipole subtraction terms are
generated with the automated Catani--Seymour procedure~\cite{Catani:1996vz}
as implemented in \tsc{Sherpa}~\cite{Gleisberg:2007md}. The code for
the evaluation of the virtual corrections has been generated by
\tsc{GoSam}~\cite{Cullen:2011ac} and is linked to \tsc{Sherpa} via the
Binoth--Les-Houches interface \cite{Binoth:2010xt,Alioli:2013nda}.
\tsc{GoSam} is an automated one-loop amplitude package, combining
automatized diagram generation and algebraic manipulation
\cite{Nogueira:1991ex,Vermaseren:2000nd, Reiter:2009ts, Cullen:2010jv}
with $d$-dimensional integrand-level reduction as implemented in the
libraries \tsc{Samurai}~\cite{Mastrolia:2010nb,vanDeurzen:2013pja}
and \tsc{Ninja}~\cite{Mastrolia:2012bu}.
Alternatively, the integrand reduction can also proceed via a
tensorial decomposition \cite{Heinrich:2010ax} using the library
\tsc{golem95C}~\cite{Binoth:2008uq,Cullen:2011kv,Guillet:2013msa}. 
In certain cases, the latter serves as the rescue route for phase
space points yielding insufficient one-loop amplitude precision in the
first place.

\begin{figure}[t!]
  \centering
  \begin{subfigure}[b]{0.3\textwidth}
    \centering
    \includegraphics[width=1\textwidth]{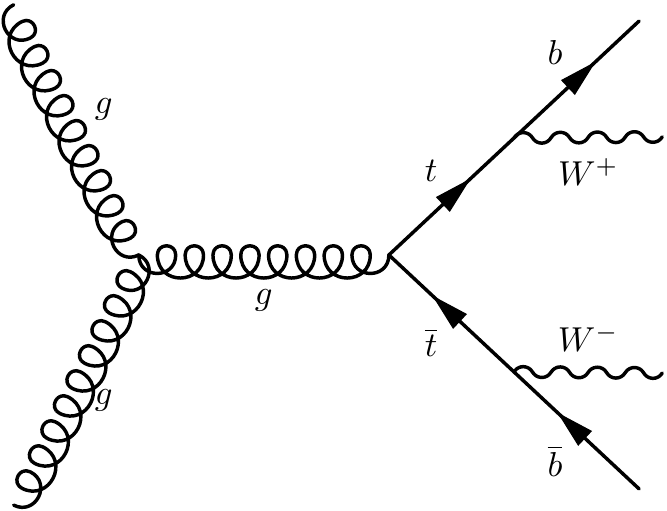}
    \caption{\label{sfig:res}}
  \end{subfigure}
  \hskip5mm
  \begin{subfigure}[b]{0.3\textwidth}
    \centering
    \includegraphics[width=1\textwidth]{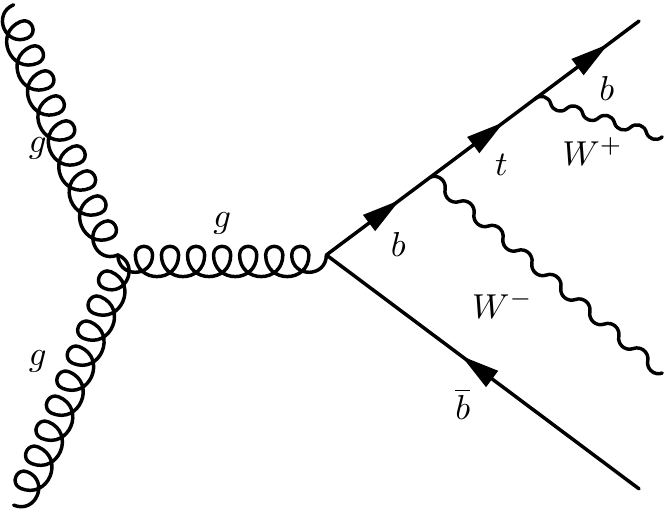}
    \caption{\label{sfig:sres}}
  \end{subfigure}
  \hskip5mm
  \begin{subfigure}[b]{0.3\textwidth}
    \centering
    \includegraphics[width=1\textwidth]{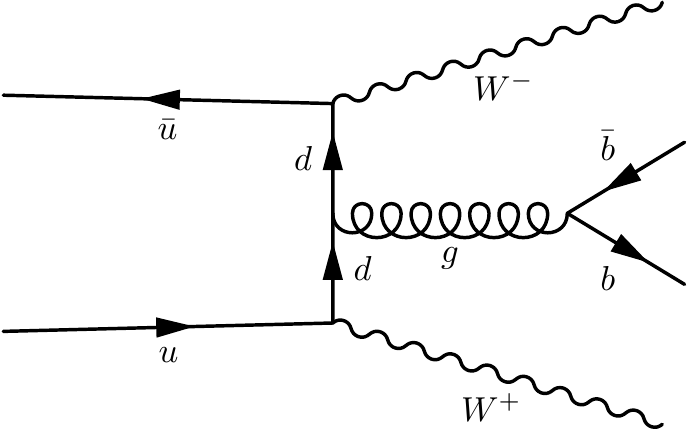}
    \caption{\label{sfig:nres}}
  \end{subfigure}
  \caption{\label{fig:resonant}
    Representative tree-level Feynman diagrams for resonant
    (\ref{sfig:res}), singly resonant (\ref{sfig:sres}) and
    non-resonant (\ref{sfig:nres}) contributions.}
\end{figure}

Our NLO accurate calculations of the $2\to4$ processes
$pp$ and $p\bar p\rightarrow W^+W^-b\bar b\rightarrow
(e^+ \nu_e)\,(\mu^- \bar{\nu}_{\mu})\,b\bar b$ provide a full
description of the final state, which is typically used as a signature
for the decay of a $t\bar t$ pair with leptonic $W$ boson decays. As
mentioned in the introduction, we include singly resonant top quark
and non-resonant contributions, see Figure~\ref{fig:resonant}.
%The latter  correspond to Feynman diagrams containing only one
%respectively no top quark propagators which can go on-shell (see
%Fig.\,\ref{fig:resonant}).
Owing to their small overall effect, diagrams that involve Higgs
bosons have been neglected throughout. Our computation relies on the
5-flavour scheme. While the subprocesses with charm and strange quarks
in the initial state are equivalent to those of the $u\bar u$ and
$d\bar d$ channels, the $b\bar b$ subprocess has to be generated
separately because the initial state $b$ quarks can propagate to the
final state, thus leading to additional diagrams.

To take the top quark decay width into account in a gauge invariant
way, the complex mass scheme~\cite{Denner:2006ic} is used. In our
setup, this amounts to replacing the top quark mass everywhere by a
complex number $\mu_t$ according to
\begin{equation}
\mu_t^2\;=\;m_t^2-i\;\!m_t\mrm{\Gamma}_t~.
\label{eq:cms}
\end{equation}
The weak mixing angle remains real-valued in our calculation, as we
neglect non-resonant $W$ and $Z$ boson contributions.

Using this setup, the correctness of the virtual amplitude has been
checked by comparing it with the results of \cite{Denner:2012yc}
for a given phase-space point. Furthermore, the calculation
of the real radiation component was verified by evaluating the cross
section for different values of the dipole $\alpha$-parameter
\cite{Nagy:1998bb}. Employing $\alpha_\trm{dp}=\{0.1,\,0.05,\,0.01\}$,
the results were found to be in agreement within the numerical
uncertainty.

\subsection{Treatment of top quarks}\label{sec:toptreat}

To investigate top quark finite-width effects, we compare the outcomes
of two different types of calculations for the
$W^+(e^+\nu_e)\,W^-(\mu^-\bar\nu_\mu)\,b\bar b$ final states:
\bi
\item[\textbf{(I)}] the {\em full or $WWb\bar b$ approach} based on
  the NLO or LO treatment of the $2\to4$ processes where finite width
  effects of the top quarks and non-resonant contributions are {\em
  fully}\/ taken into account, and
\item[\textbf{(II)}] the {\em factorized or $t\bar t$ approach}\/
  based on the narrow width approximation where the production of the
  top quarks {\em factorizes}\/ from their decays. In our case, the
  higher-order treatment will be limited to the production part: all
  NLO corrections {\em only}\/ apply to the $2\to2$ processes of top
  quark pair production, i.e.~the top quark decays are still described
  with leading order accuracy. In the NLO context, one commonly refers
  to this approach as ``narrow width approximation with leading-order
  decays'' to distinguish it from the complete NLO treatment combining
  $t\bar t$ production and decay in the narrow width approximation, as
  accomplished in Ref.~\cite{Melnikov:2009dn}.
\ei

\begin{figure}[t!]
  \centering
  \begin{subfigure}[b]{0.45\textwidth}
    \centering
    \includegraphics[width=1\textwidth]{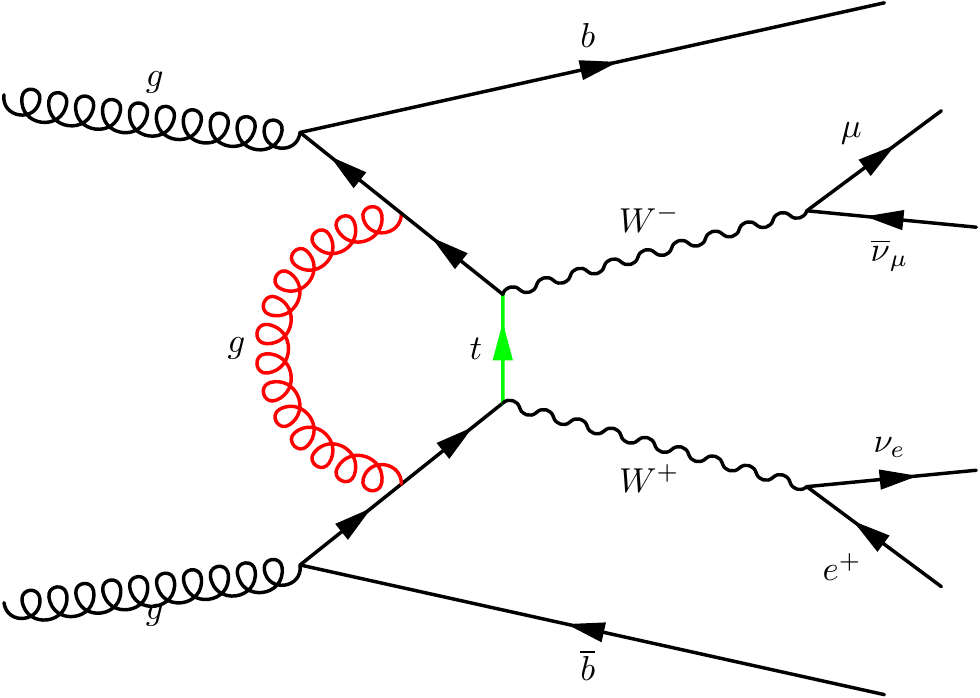}
    \caption{\label{sfig:nresloop}}
  \end{subfigure}
  \hskip10mm
  \begin{subfigure}[b]{0.45\textwidth}
    \centering
    \includegraphics[width=1\textwidth]{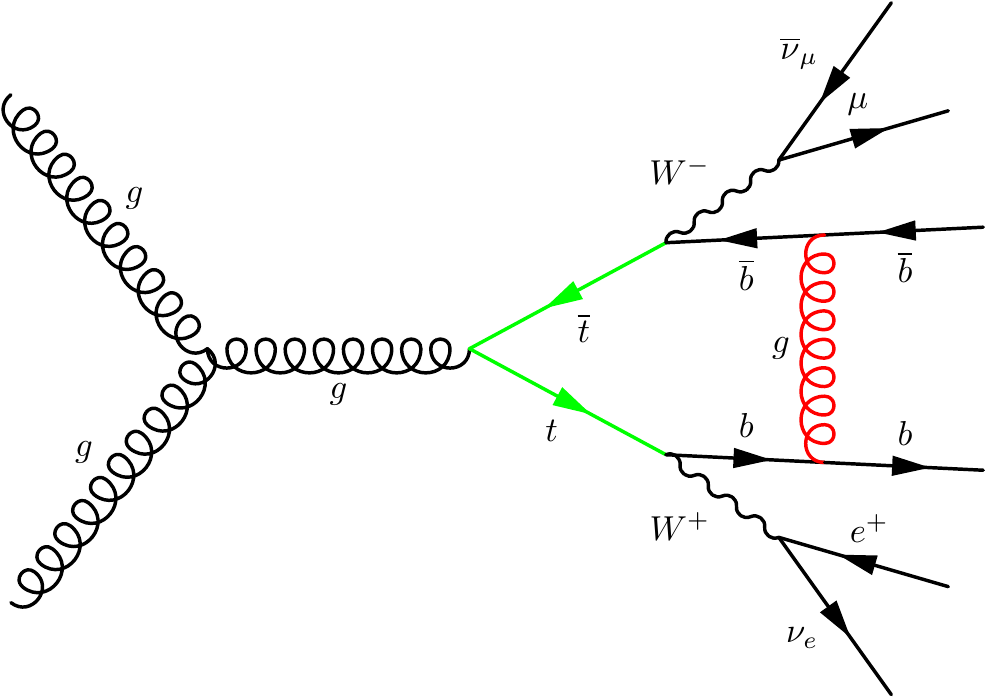}
    \caption{\label{sfig:nfacloop}}
  \end{subfigure}
  \caption{\label{fig:diagloop}
    Examples of one-loop Feynman diagrams contributing to the full
    calculation: a non-resonant diagram (\ref{sfig:nresloop}) and a
    non-factorizable virtual contribution (\ref{sfig:nfacloop}).}
\end{figure}

The narrow width approximation (NWA) is motivated by the fact that, in
the limit $\mrm{\Gamma}_t\to0$, the denominator of the top quark
propagator can be written as
\begin{equation}
\lim_{\mrm{\Gamma}_t\to0}\frac{1}{(p_t^2-m_t^2)^2+m_t^2\mrm{\Gamma}_t^2}\;=\;
\frac{\pi}{m_t\mrm{\Gamma}_t}\,\delta(p_t^2-m_t^2) +
\mathcal{O}\left(\frac{\mrm{\Gamma}_t}{m_t}\right)~.
\label{eq:nwa}
\end{equation}
%The $\delta$ distribution sets the top quark on its mass-shell. 
Since this approximation introduces a factor of $1/\mrm{\Gamma}_t$ for
each top quark resonance, singly resonant and non-resonant
contributions are suppressed in the $\mrm{\Gamma}_t\to0$ limit.
Consequently, one only keeps the Feynman diagrams where two top quarks
can become resonant, because only those are proportional to
$1/\mrm{\Gamma}_t^2$. In the $\mrm{\Gamma}_t\to0$ limit, the full
process therefore factorizes into top quark pair production and decay,
i.e.\ $pp$ and $p\bar p\to t\bar t\rightarrow W^+b\;W^-\bar{b}$.
%\begin{equation}
%\text{d}\sigma_{\text{NWA}}^{\text{LO}}=\frac{1}{(\Gamma_t^{\text{LO}})^2}\left[\text{d}\sigma_{t\bar{t}}^{\text{LO}}\text{d}\Gamma_{t\rightarrow W^+b}^{\text{LO}}\text{d}\Gamma_{\bar{t}\rightarrow W^-\bar{b}}^{\text{LO}}\right].
%\label{eq:factlo}
%\end{equation}
%Here the notation of \cite{Denner:2012yc} is used. 
%At NLO, corrections to both $t\bar{t}$ production and decay have to
%be taken into account.
%\begin{equation}
%\text{d}\sigma_{\text{NWA}}^{\text{NLO}}=\frac{1}{(\Gamma_t^{\text{NLO}})^2}\left[(\text{d}\sigma_{t\bar{t}}^{0}+\text{d}\sigma_{t\bar{t}}^{1})\text{d}\Gamma_{t}^{0}\text{d}\Gamma_{\bar{t}}^{0}+\text{d}\sigma_{t\bar{t}}^{0}(\text{d}\Gamma_{t}^{1}\text{d}\Gamma_{\bar{t}}^{0}+\text{d}\Gamma_{t}^{0}\text{d}\Gamma_{\bar{t}}^{1})\right].
%\label{eq:factnlo}
%\end{equation}
Thus, when working in the NWA, at NLO one also neglects radiative
corrections that either connect production and decay, or both decays.
Two example Feynman diagrams contributing to the virtual corrections,
which are not present in the NWA, are given in Figure~\ref{fig:diagloop}.

From Eq.~\eqref{eq:nwa} one recalls that the contributions neglected
in the NWA are suppressed by powers of $\mrm{\Gamma}_t/m_t\lesssim1\%$.
While this is true for sufficiently inclusive observables, the
corrections can be much larger for observables such as \mlb, the
system invariant mass of the charged lepton and the (associated) $b$
jet. We will discuss this issue in more detail in Section \ref{sec:mlb}.

\subsection{General input parameters}

For the (N)LO calculations, the MSTW2008(N)LO parton distributions
\cite{Martin:2009iq} were used, relying on the strong coupling
constant, $\alpha_\mrm{s}$, and its running as provided by these PDF
parametrizations. The electroweak parameters are given in the
$G_{\mu}$ scheme:
\begin{equation}
%\begin{aligned}
%G_{\mu} &=1.16637\cdot10^{-5}\,\text{GeV}^{-2} \\
%M_{W} &=80.399\text{GeV} \\
%M_{Z} &=91.1876\text{GeV}
%\end{aligned}
  \begin{matrix}
   G_{\mu} &=& ~1.16637\cdot10^{-5}&\gev^{-2},~~& \\[1mm]
   M_{W}  &=& 80.399\gev\,,\hphantom{0}     && \mrm{\Gamma}_W &=& 2.0997\gev\,,\\
   M_{Z}  &=& 91.1876\gev\,,                && \mrm{\Gamma}_Z &=& 2.5097\gev\,.
  \end{matrix}
  \label{eq:ewparam}
\end{equation}
%, which corresponds to setting its mass to infinity.
All quarks other than the top quark are taken to be massless. For the
top quark mass, we use $m_t=172.0\gev$. From the parameters given
above, it is possible to derive the value of the top quark decay width
at LO and NLO using the expressions calculated in \cite{Jezabek:1988iv}.
We use the numerical values
\begin{equation}
  \begin{aligned}
    \mrm{\Gamma}_t^\srm{LO} &\;=\;1.4426\gev\,,\\
    \mrm{\Gamma}_t^\srm{NLO} &\;=\;1.3167\gev\;.
  \end{aligned}
  \label{eq:topidth}
\end{equation}

\boldmath
\subsection{Numerical results for LHC collisions at $7\tev$}
\unboldmath\label{sec:res}

Using the full approach, cf.~Section~\ref{sec:toptreat}~(I), we now
study the impact of the NLO corrections to $W^+W^-b\bar b$ production
in dilepton final states at the LHC for a collision energy of $7\tev$.
To produce these results, we impose the following set of kinematical
requirements: all final state partons are clustered into jets with a
separation in azimuthal angle ($\phi$) and pseudo-rapidity ($\eta$)
space defined by
\begin{equation}
  \upDelta R\;=\;\sqrt{\upDelta\phi^2 + \upDelta\eta^2}\;>\;0.5~,
  \label{eq:deltarj}
\end{equation}
using the anti-$k_T$ jet algorithm \cite{Cacciari:2005hq,Cacciari:2008gp}
implemented in \tsc{FastJet} \cite{Cacciari:2011ma}. Each event is
required to contain at least two $b$ jets obeying the conditions
\begin{equation}
  \begin{matrix}
    p_{T,b}\;>30\gev &\text{\quad and\quad}& |\eta_{b}|\;<\;2.5~.
  \end{matrix}
  \label{eq:bcuts}
\end{equation}
The requirements on the charged leptons ($l$) and the missing energy
are\,\footnote{Here, we employ the transverse vector sum of the
  neutrinos to determine the missing energy.}
\begin{equation}
  \begin{matrix}
    p_{T,l}\;>\;20\gev\,, &\quad& |\eta_{l}|\;<\;2.5
    &\text{\quad and\quad}& \slashed{p}_T\;>\;20\gev\,,
  \end{matrix}
  \label{eq:lcuts}
\end{equation}
respectively.

\begin{figure}[t!]
  \centering
  \begin{subfigure}[]{0.49\textwidth}
    \centering
    \includegraphics[width=1.0\textwidth]{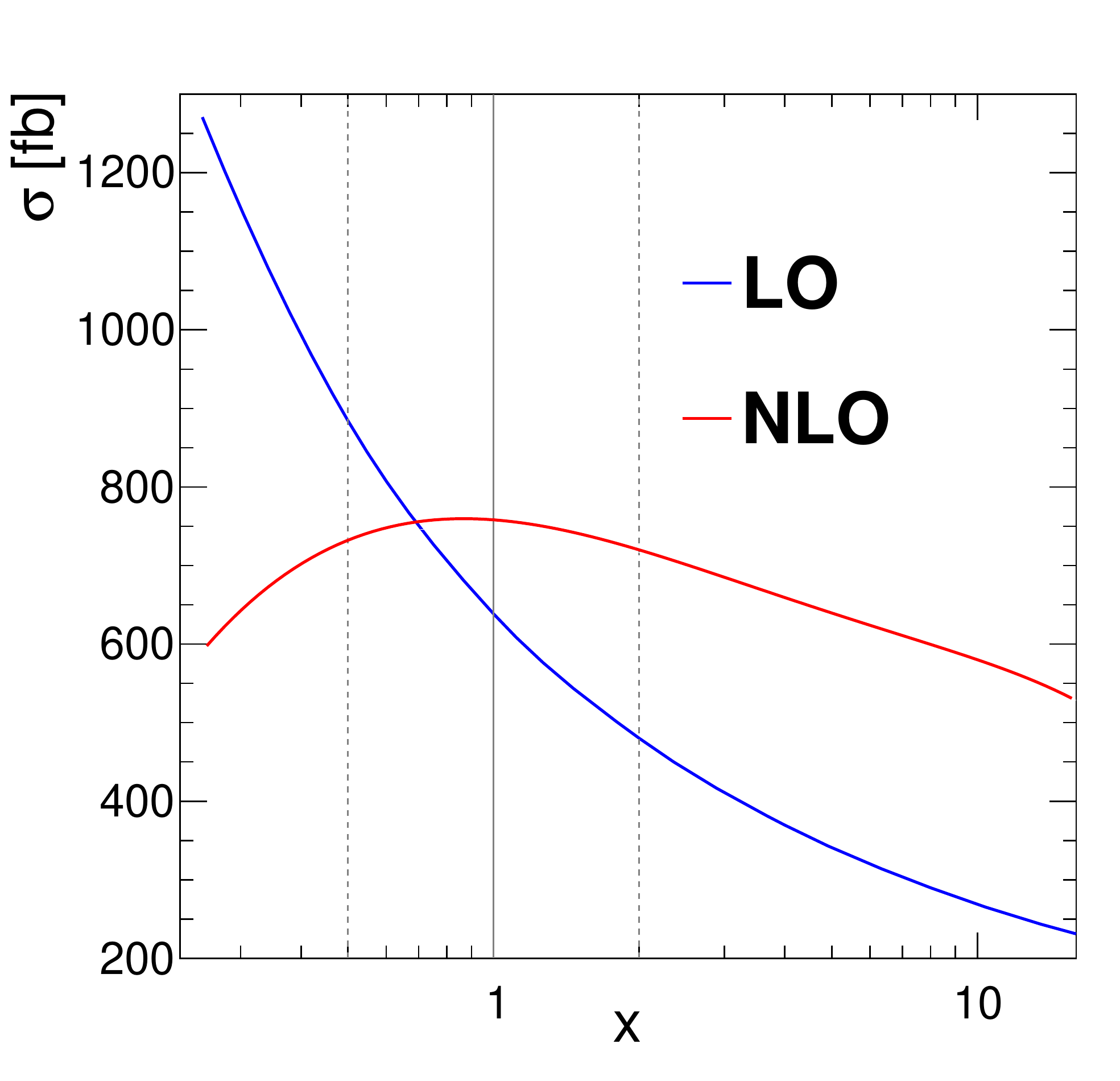}
    \caption{\label{sfig:scale}}
  \end{subfigure}
  \hskip1mm
  \begin{subfigure}[c]{0.49\textwidth}
    \centering
    \includegraphics[width=0.99\textwidth]{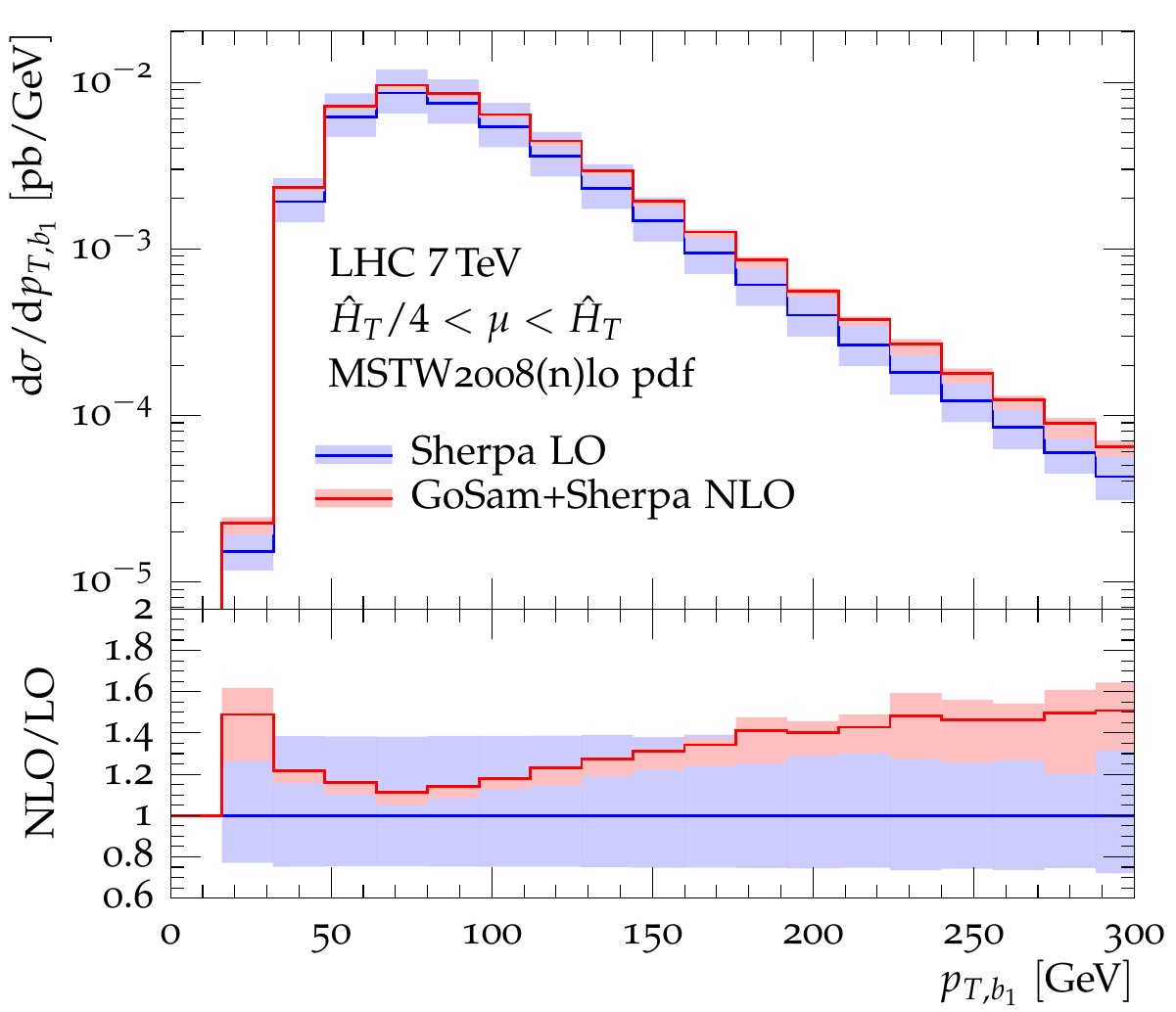}
    \caption{\label{sfig:ptbmax}}
  \end{subfigure}
  \caption{\label{fig:misc}
    Scale variation of the LO and NLO cross sections in the full
    approach (\ref{sfig:scale}), ranging from $x=1/4$ to $x=16$ where
    $x=2\,\mu/\hat H_T$ and $\mu=\mu_\srm{R}=\mu_\srm{F}$.
    Transverse momentum distribution of the leading $b$ jet at LO and
    NLO in the full approach (\ref{sfig:ptbmax}). The bands were
    obtained by varying $\mu$ by a factor of two around the central
    scale $\hat H_T/2$.}
\end{figure}

Similarly to the computation presented in Ref.~\cite{Frederix:2013gra}, 
we use $\hat H_T$, specified through
\begin{equation}
  \hat H_T\;=\;\sum_{i}p_{T,i}~,
\label{eq:ht}
\end{equation}
to define our default scale choice, $\mu_\srm{R}=\mu_\srm{F}=\mu\equiv\hat H_T/2$,
in setting the renormalization and factorization scales. Note that the
sum in Eq.~\eqref{eq:ht} runs over all final state particles,
including the neutrino momenta. The central scale is chosen to be
$\mu\equiv\hat H_T/2$ because this helps minimize the difference
between the LO and the NLO cross section as well as the uncertainty
induced by scale variations. Concerning the latter fact, this scale
uncertainty estimate turns out to be smaller than those obtained from
the scale choices $\mu=\hat H_T$ or $\mu=\mt$.

\bigskip
Given the settings above, we obtain the LO and NLO inclusive cross
sections for the full approach, reading
\begin{equation}
  \begin{aligned}
    \sigma_\srm{LO}\;\mrm{[fb]} &\;=\;
    638.4\,^{+38.5\%}_{-24.8\%}\,(\text{scale})\;\pm\;0.03\%\,(\text{stat})~,\\[1mm]
    \sigma_\srm{NLO}\;\mrm{[fb]} &\;=\;
    758.5\,^{-2.5\%}_{-5.3\%}\,(\text{scale})\;\pm\;0.2\%\,(\text{stat})~.
  \end{aligned}
  \label{eq:xs}
\end{equation}
This corresponds to a $K$-factor of about $1.2$. The scale
uncertainties given in Eqs.~\eqref{eq:xs} are obtained from varying
the central scale $\hat H_T/2$ by factors of two, i.e.~using the
multiplicative factors $x=1/2$ and $x=2$ where $x=\frac{\mu}{\hat H_T/2}$.
We also indicate the statistical uncertainty of our Monte Carlo
integrations. As the NLO cross section at the central scale is larger
than for both the upwards and downwards scale variation (see
Figure~\ref{sfig:scale}), the NLO scale uncertainties given in
Eqs.~\eqref{eq:xs} can only be negative. Another, perhaps more
reasonable way of estimating the scale uncertainties is to consider
the highest and lowest cross section within a $\mu$ range determined
by factors of two around the central scale. According to this
procedure, the maximum (minimum) value for the NLO cross section is
$759.6\mrm{\:fb}$ ($718.3\mrm{\:fb}$) occurring at a scale slightly
lower than (twice as large as) the central one. Finally, and for an
extended $x$ range, the sensitivity of the LO and NLO cross sections
to the choice of the renormalization and factorization scales,
$\mu_\srm{R}=\mu_\srm{F}$, is shown in Figure~\ref{sfig:scale}. Here,
the scales have been varied between $\hat H_T/8$ and $8\,\hat H_T$
retaining $\mu_\srm{R}=\mu_\srm{F}$.

In Figure~\ref{sfig:ptbmax}, we now present a first differential
distribution comparing the LO and NLO predictions with their absolute
normalizations for the leading $b$ jet transverse momentum. The
respective bands have been determined, as before, from scale
variations evaluated at $x=1/2$ and $x=2$. The ``NLO/LO'' ratio
plot clearly exhibits the reduction of the theory uncertainties, as
well as the hardening of the $p_T$ spectrum owing to the generation of
real radiation that recoils against the $t\bar t$ system. We
furthermore notice that shape changes to the $p_{T,b_1}$ distribution
only occur at NLO; at LO, the shape is more or less predicted to be
constant, as reflected by the uniform envelope around the $p_{T,b_1}$
LO prediction.

%We also highlight the differences between the full NLO calculation
%and the calculation where only the $t\bar{t}$ production cross
%section is calculated at NLO, combined with the decays keeping the
%full spin correlations.

\begin{figure}[t!]
  \centering
  \begin{subfigure}[b]{0.49\textwidth}
    \centering
    \includegraphics[width=1.0\textwidth]{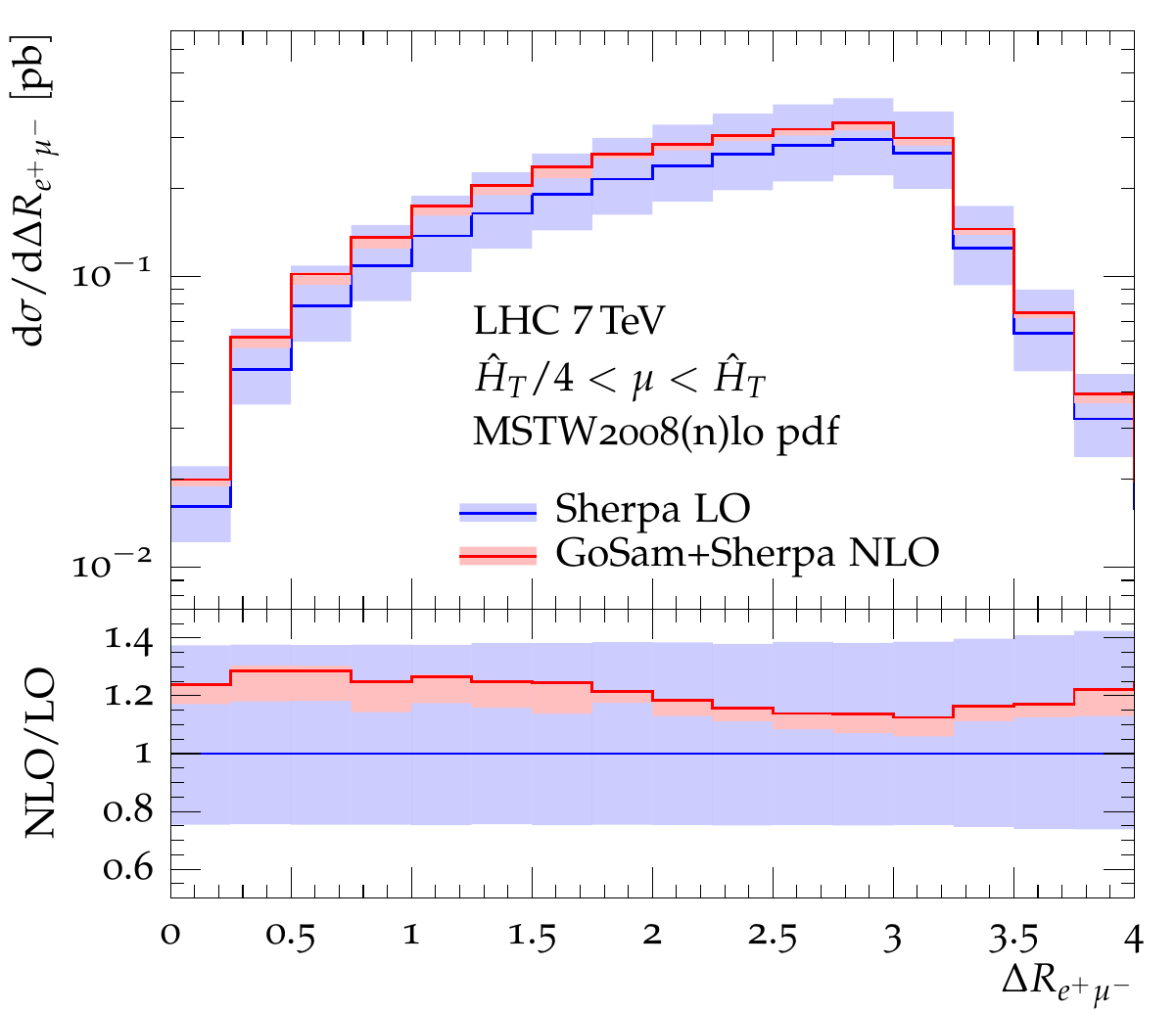}
    \caption{\label{sfig:rll}}
  \end{subfigure}
  \hskip1mm
  \begin{subfigure}[b]{0.49\textwidth}
    \centering
    \includegraphics[width=1.0\textwidth]{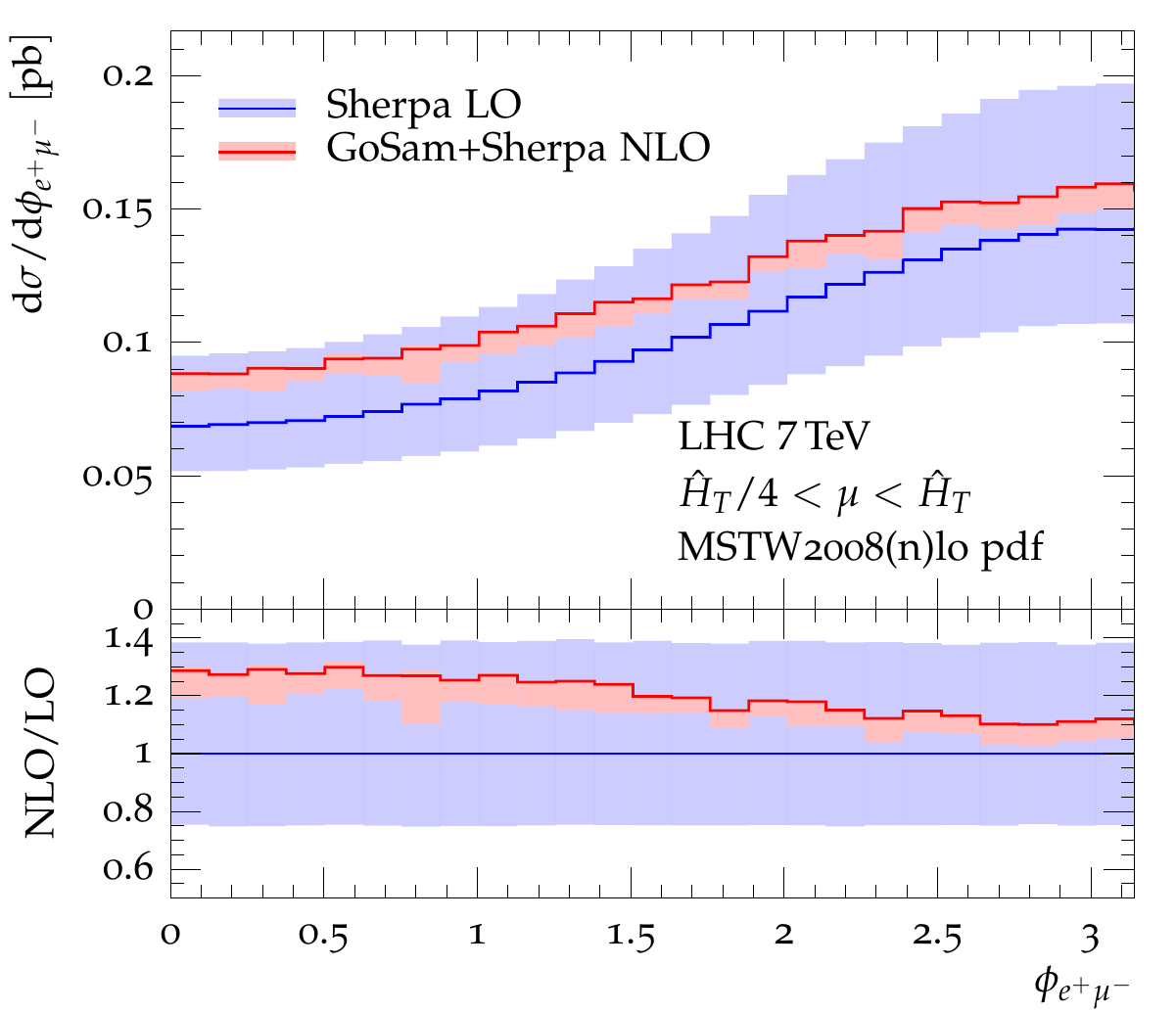}
    \caption{\label{sfig:phi}}
  \end{subfigure}
  \caption{\label{fig:angle}
    Differential distributions of the $\upDelta R$ separation
    (\ref{sfig:rll}) and the relative azimuthal angle between the two
    charged leptons (\ref{sfig:phi}) in $W^+W^-b\bar b$ production at
    LO and NLO (in the full approach). The bands were obtained by
    varying the scales by a factor of two around the central scale
    $\hat H_T/2$.}
\end{figure}

Figure~\ref{fig:angle} shows differential distributions related to the
two charged leptons stemming from the $W$ boson decays.
Figure~\ref{sfig:rll} displays the $\upDelta R$ separation between the
leptons, $e^+$ and $\mu^-$, while Figure~\ref{sfig:phi} shows the
projection of the relative angle between these two leptons onto the
plane transverse to the beam axis, $\phi_{e^+\mu^-}$. These lepton
correlations play an important role in the measurement of top quark
spin correlations at the LHC. For both distributions, we observe a
substantial reduction of the scale uncertainties at NLO. Again, scale
variations by and large do not affect the LO shapes, a description
at NLO therefore is much more reliable. The distribution of the
azimuthal angle $\phi_{e^+\mu^-}$ receives the largest NLO corrections of
$\mathcal{O}(30\%)$ in regions where the separation between the two
leptons is small. Even for small angles, the $K$-factor varies no more
than $\sim10\%$.

\begin{figure}[t!]
  \centering
  \begin{subfigure}[b]{0.49\textwidth}
    \centering
    \includegraphics[width=1\textwidth]{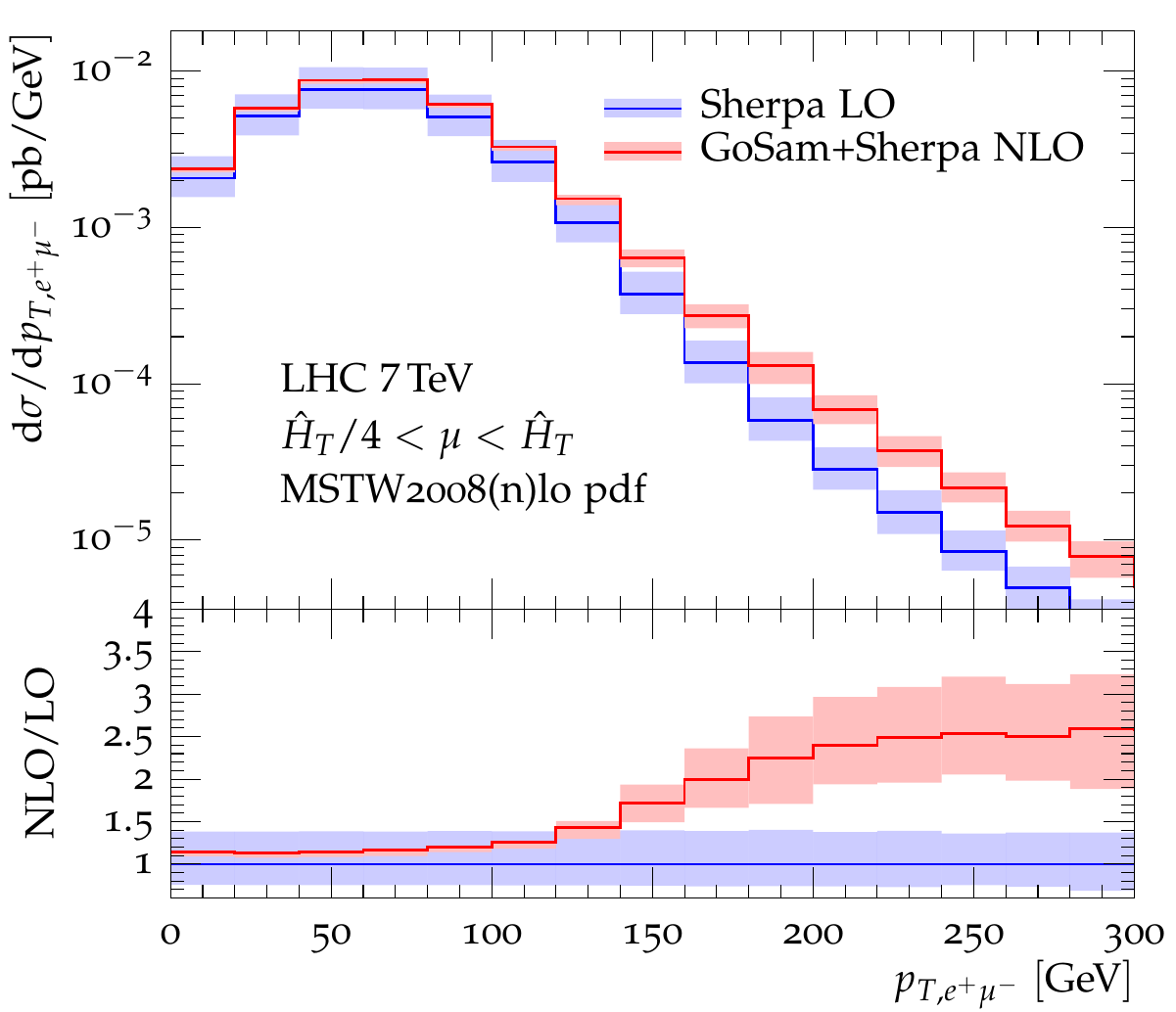}
    \caption{\label{sfig:ptll}}
  \end{subfigure}
  \hskip1mm
  \begin{subfigure}[b]{0.49\textwidth}
    \centering
    \includegraphics[width=1\textwidth]{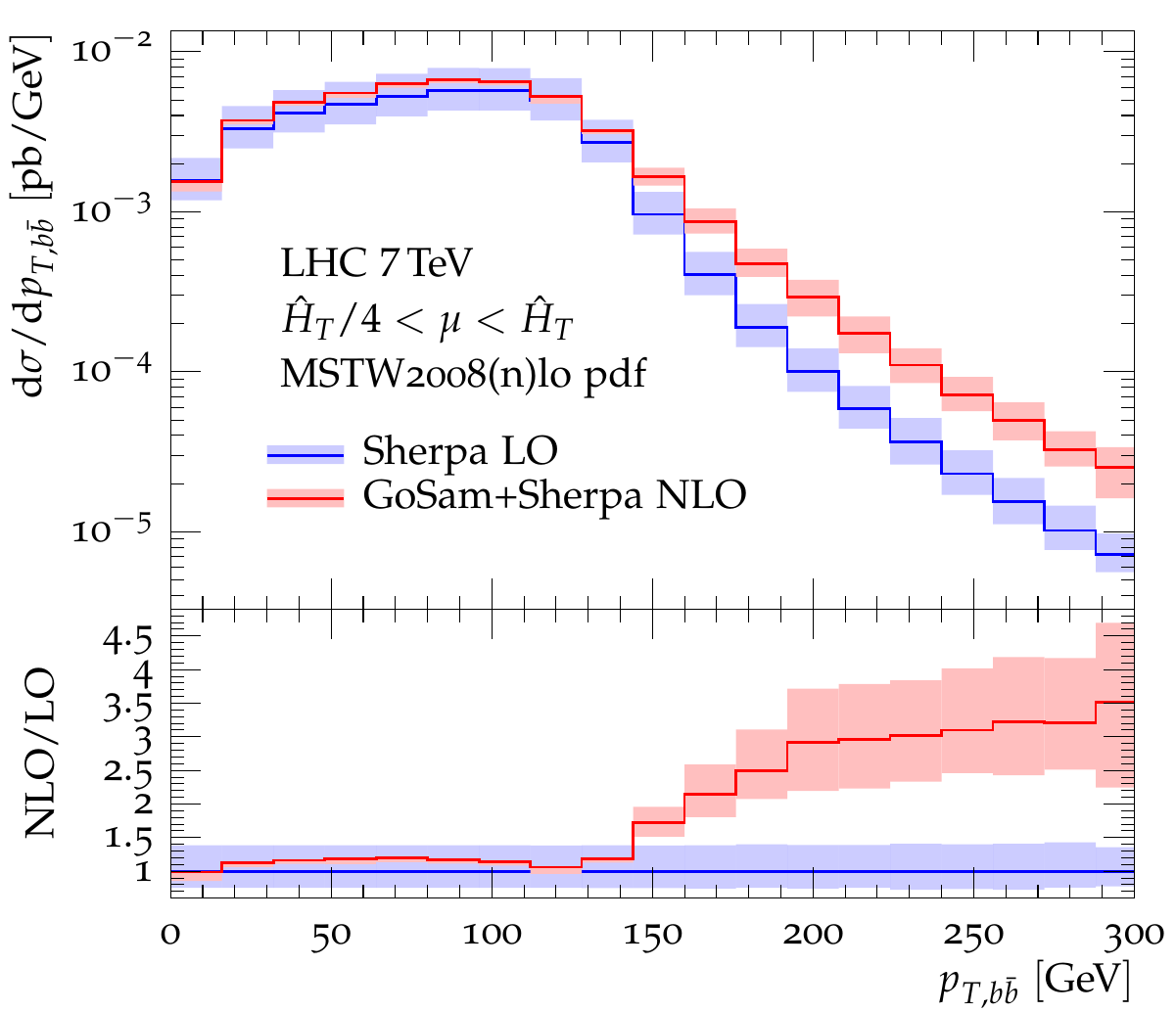}
    \caption{\label{sfig:ptbb}}
  \end{subfigure}
  \caption{\label{fig:pttwo}
    Transverse momentum spectra of (\ref{sfig:ptll}) the charged
    lepton pair and (\ref{sfig:ptbb}) the $b\bar b$ system, i.e.~the
    system consisting of the two leading $b$ jets, in $W^+W^-b\bar b$
    production at LO and NLO (in the full approach). The bands were
    obtained by varying the scales by a factor of two around the
    central scale $\hat H_T/2$.}
\end{figure} 

Figures~\ref{sfig:ptll} and \ref{sfig:ptbb} visualize the system
transverse momentum spectra of the two charged leptons and the two
$b$ jets, respectively. These observables receive large NLO
corrections with $K$-factors growing as large as $\sim3$ in the region
of hard $p_T$. The reason lies in the generation of the real radiation
component that recoils against the entire $WWb\bar b$ system. This
component, which is absent at LO, leads to a $p_T$ imbalance between
the $WW$ and $b\bar b$ subsystems, which can be noticed in particular
for $p_{T,b\bar b}\gtrsim2\,M_W$ where the effect becomes largest.
Therefore, it is not surprising that the scale uncertainty band
associated with the LO distribution does not contain the NLO result in
the tail of this and similar distributions. For the charged lepton
pair $p_T$, the effect is somewhat washed out and smaller -- simply
because the dileptons do not carry the full information on the $p_T$
of the $WW$ system.

%%%%%%%%%%%%%%%%%%%%%%%%%%%%%%%%%%%%%%%%%%%%%%%%%%

\section{Phenomenological studies}\label{sec:pheno}
%\subsection{Top quark mass measurements}%\label{sec:mt}
%\input{topmass_old}
 In the following we will concentrate on two applications of our parton level
 calculations using both the full and the factorized approach, as described in
 Section~\ref{sec:toptreat}.
 Firstly, any shape-based $m_t$ measurement relies on the precise modelling of
 the differential distribution whose shape depends on the value of the top quark
 mass. Shape uncertainties induced by $\mu_\srm{F,R}$ scale variations will
 therefore impact the accuracy of $m_t$ measurements. For the example of the
 \mt measurement based on the \mlb\ observable, we will study this issue in
 detail.
 Secondly, in the context of $t\bar t$ asymmetry measurements, it is crucial to
 understand the relation between top quark and lepton-based asymmetries. We will
 discuss the strength of their correlation, particularly for Tevatron analyses
 using the dilepton channel.

\subsection{Top quark mass measurements}

 As the top quark mass \mt\ is not a physical observable, its definition is
 scheme dependent.  The most commonly used mass definitions are the pole mass
 and the \MSbar\ mass. The different masses are related by a perturbative
 series, see e.g.~Refs.~\cite{Chetyrkin:1999ys,Melnikov:2000qh}.

 The pole mass scheme is a long distance scheme, where implicitly the top quark
 is considered as a stable particle, with the pole mass being defined as the
 real part of the pole of the propagator. However, the fact that quarks do not
 appear as isolated particles implies that non-perturbatively there is no pole
 in the scattering amplitude due to the quark propagator, and only in
 perturbation theory the pole mass is properly defined.  The pole mass also gets
 corrections of order $\mrm{\Lambda}_\srm{QCD}$ from the infinite sum of self
 energy insertions, which is called the renormalon
 ambiguity~\cite{Beneke:1994sw,Beneke:1998ui}.

 The most common short distance scheme is the modified minimal subtraction
 (\MSbar) scheme. In contrast to the pole mass, the \MSbar\ mass is not
 sensitive to corrections related to the renormalon ambiguity.  Despite this
 fact, a similar convergence behaviour of the top quark mass in both schemes has
 been observed up to NNLO~\cite{Ahrens:2011px,Mangano_top2013}.

 Experimental results for the top quark mass are obtained by comparing
 experimental observables to the prediction from Monte Carlo event
 generators. The exact relation between the mass parameter $m_t^\srm{MC}$ used
 in the Monte Carlo program and the pole mass at a given order in perturbation
 theory is still an open
 issue~\cite{Buckley:2011ms,Hoang:2008xm,Agashe:2013hma,Juste:2013dsa}. The
 related uncertainty is estimated to be about $1\gev$.  A study aiming at
 disentangling systematically genuine non-perturbative effects from perturbative
 ones can be found in Ref.~\cite{Skands:2007zg}, see
 also~\cite{Mangano_top2013}.

 To avoid these problems, it has been suggested to determine the \MSbar\ mass by
 comparing the measured total cross section for top quark pair production with a
 fixed order calculation performed in the \MSbar\
 scheme~\cite{Abazov:2009ae,Langenfeld:2009wd}.  However, this method cannot
 circumvent the problem completely, as the experimental determination of the
 $t\bar{t}$ cross section also has to rely on
 $m_t^\srm{MC}$~\cite{Nisius:2012gm}.

\boldmath
\subsection{Mass determination using the $\mlb$ observable}
\unboldmath\label{sec:mlb}
 An observable which has recently been used for a top quark mass determination
 at the LHC~\cite{ATLAS-CONF-2013-077} is the invariant mass of a charged lepton
 and a $b$ jet, $\mlb^2=(p_l+p_b)^2$, where $p_b$ denotes the four-momentum of
 the $b$ jet.
 Already in Ref.~\cite{Biswas:2010sa} this observable has been studied in view
 of top quark mass determinations, however, non-factorizing contributions have
 not been taken into account by that calculation.

%------------------------------------------------------------------
\begin{figure}[t!]
  \centering
  \begin{subfigure}[b]{0.49\textwidth}
    \centering
    \includegraphics[width=1\textwidth]{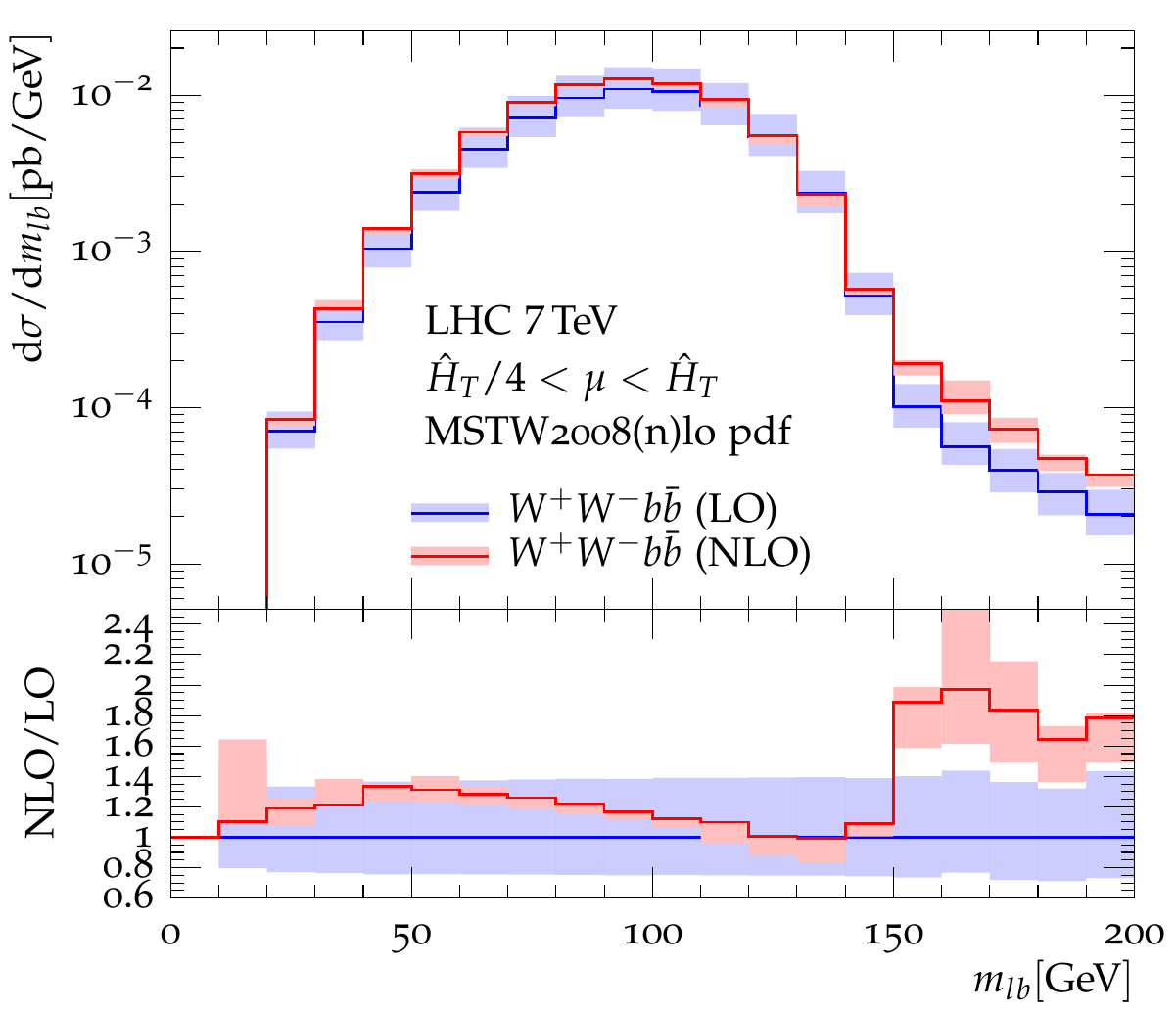}
    \caption{\label{sfig:mlb2}}
  \end{subfigure}
  \hskip1mm
  \begin{subfigure}[b]{0.49\textwidth}
    \centering
    \includegraphics[width=1\textwidth]{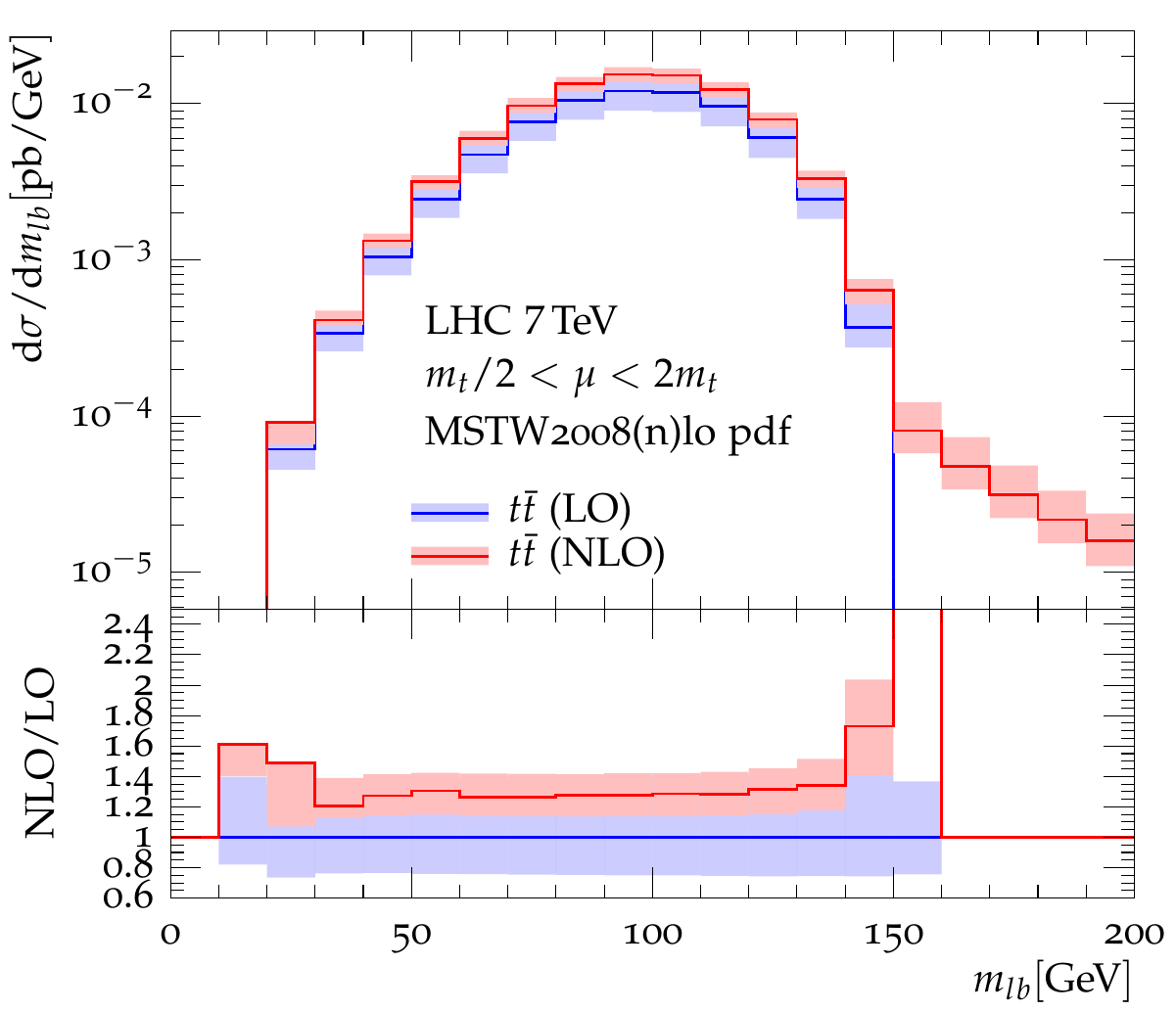}
    \caption{\label{sfig:mlbnwa}}
  \end{subfigure}
  \caption{\label{fig:mlbnlo}
    Distribution of \mlb\ at LO (blue lines) and NLO (red lines),
    including standard scale variations for (\ref{sfig:mlb2}) the full
    calculation, and (\ref{sfig:mlbnwa}) the factorized calculation. In
    addition shown are the ratios using the respective LO central
    predictions as their references.
    The scale choices differ: they are $\mu=\hat{H}_T/2$ and $\mu=m_t=172.5\gev$
    for the full and the factorized approach, respectively.}
\end{figure}
%------------------------------------------------------------------

 The latest ATLAS result~\cite{ATLAS-CONF-2013-077} of \mt\ using the
 \mlb\ observable is:
 \be
 \mt\;=\;173.09\pm 0.64\,(\mathrm{stat})\pm 1.50\,(\mathrm{syst})~\gev~.
 \label{mtatlas}
 \ee
 The systematic uncertainty is dominated by jet energy scale uncertainties, and
 the theoretical uncertainties assigned to this measurement amount to
 about~$0.8\gev$.

 One complication arises from the fact that there are two top quarks and
 therefore two possible \mlb\ values per event. Since experimentally, the charge
 of the $b$ quark initiating a jet cannot be reconstructed on an event-by-event
 basis, one needs a criterion to identify a pair of a charged lepton
 and a $b$ jet as stemming from the same top quark decay.
 Using events generated with the \tsc{MC@NLO} Monte Carlo
 program~\cite{Frixione:2002ik}, different strategies for this assignment were
 investigated by ATLAS.
 Following the procedure given in Ref.~\cite{ATLAS-CONF-2013-077}, the
 algorithm applied here is to choose the combination, i.e.~the
 $(l^+b\text{-jet},\,l^-b\text{-jet}')$ pairing, which  minimizes the
 sum of the two \mlb\ values per event.
 Finally, the \mlb\ observable used in the analysis is the mean of the two \mlb\
 values per event obtained when applying the above procedure.

\boldmath
\subsubsection{Parton level $\mlb$ predictions at NLO}
\unboldmath

 For our calculations of the \mlb\ distribution, we follow the ATLAS procedure as
 outlined above. We use $m_t=172.5\gev$ as our default top quark mass and employ
 the ATLAS kinematic requirements for $7\tev$ LHC $pp$ collisions:
 we require exactly two oppositely charged leptons (electrons with
 $p_T>25\gev$, and muons with $p_T>20\gev$) in the pseudo-rapidity
 range $|\eta_l|<2.5$, and two $b$ jets with $p_{T,b}>25\gev$,
 $|\eta_b|<2.5$ and $\Delta R>0.4$, using the anti-$k_T$ algorithm.
 The leptons have to be isolated from the jets with $\Delta R_{l,j}>0.4$.
 Lastly, $H_T$ defined as the sum over the transverse momenta of charged leptons
 and jets has to be larger than $130\gev$.

 For the two types of calculations described in Section~\ref{sec:toptreat},
 Figure~\ref{fig:mlbnlo} shows the corresponding \mlb\ distributions
 at LO and NLO, including their respective scale variation bands as
 well as the ratios taken with respect to the central LO prediction.
 These results have been obtained from the full calculation evaluated with
 $\mu=\hat{H}_T/2$ (\ref{sfig:mlb2}), and from the factorized calculation
 evaluated at $\mu=m_t$ (\ref{sfig:mlbnwa}).

%------------------------------------------------------------------
\begin{figure}[t!]
  \centering
  \begin{subfigure}[b]{0.49\textwidth}
    \centering
    \includegraphics[width=1\textwidth]{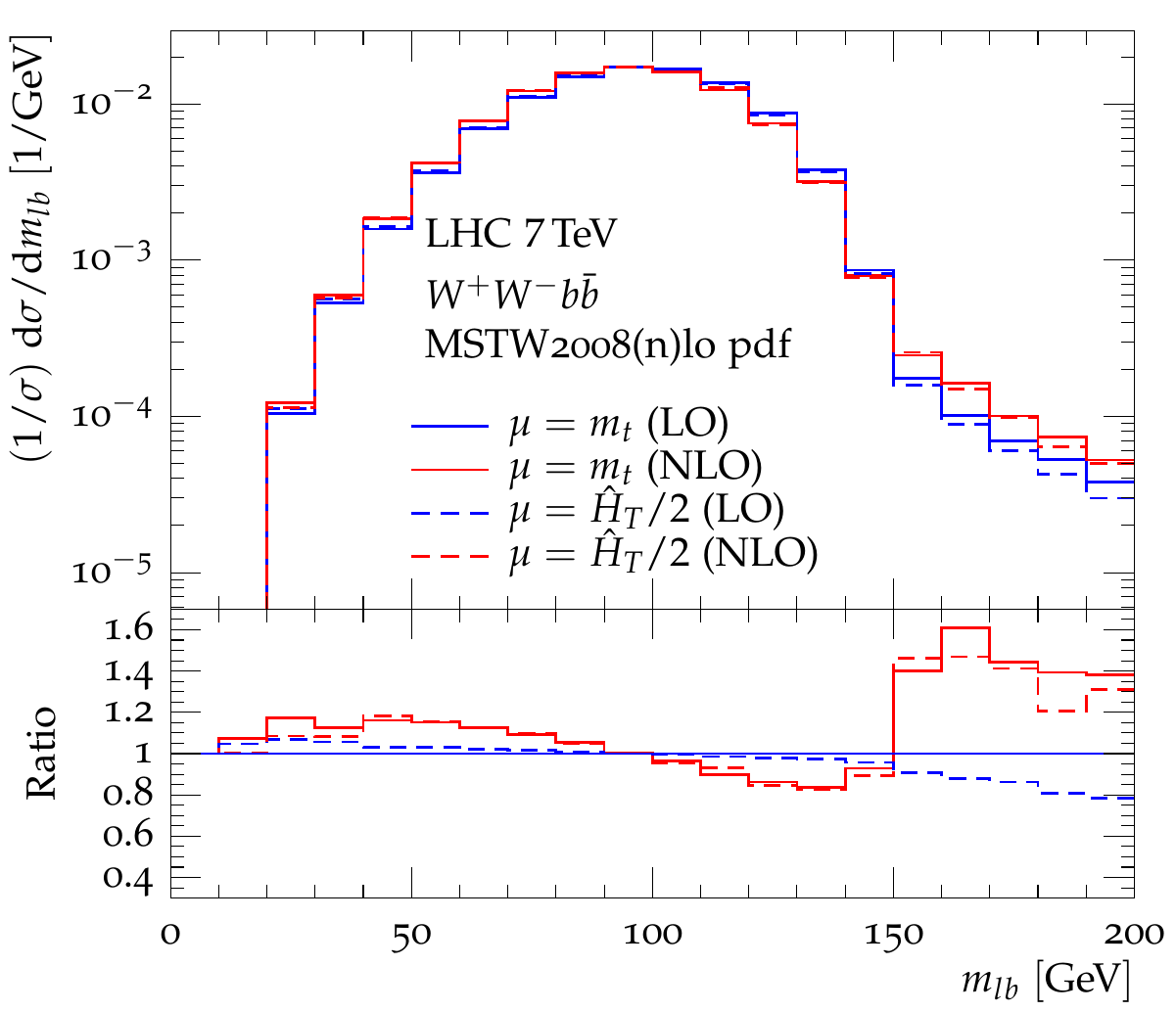}
    \caption{\label{sfig:mlbcompwwbb}}
  \end{subfigure}
  \hskip1mm
  \begin{subfigure}[b]{0.49\textwidth}
    \centering
    \includegraphics[width=1\textwidth]{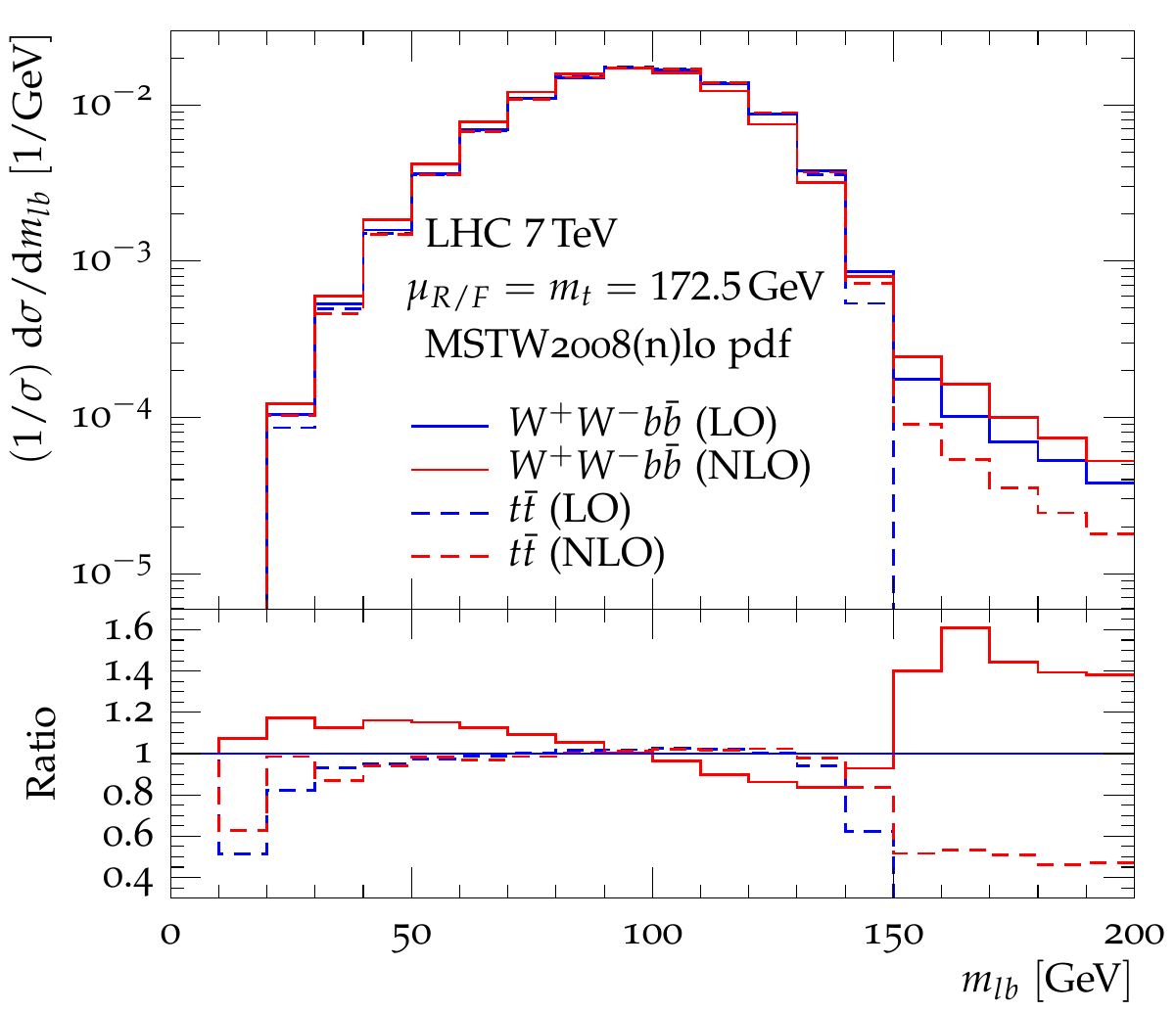}
    \caption{\label{sfig:mlbcomptt}}
  \end{subfigure}
  \caption{\label{fig:mlbcompare}
    Various normalized LO and NLO predictions of the \mlb\ distribution.
    Results (solid lines) using the full calculation and the fixed scale
    choice $\mu=m_t=172.5\gev$ are compared to (\ref{sfig:mlbcompwwbb})
    results generated with our default dynamical scale choice of
    $\mu=\hat{H}_T/2$, and (\ref{sfig:mlbcomptt}) to results of the
    factorized approach for the same fixed scale choice, $\mu=m_t$.
    The prediction of the full calculation at LO utilizing the fixed
    scale serves as the reference curve in both ratio distributions shown.}
\end{figure}
%------------------------------------------------------------------

 The full and factorized calculations exhibit a few interesting differences.
 Firstly, the uncertainty bands of the computations in the factorized approach
 are wider than those of the respective full computations.
 Secondly, the NLO corrections in the factorized approach
 (cf.~Figure~\ref{sfig:mlbnwa}) mainly affect the event rate, while
 the LO and NLO shapes of the \mlb\ distribution are very similar. The
 only exception is the region $\mlb>150\gev$, where the difference is
 caused by the fact that the LO factorized calculation has a sharp
 cut-off at $\mlb=\sqrt{\mt^2-m_W^2}$.
 Even for the full approach, the differences in the tail are found to
 exceed the estimate from scale variations of the LO calculation.
%\footnote{instead of
%``In this region, for both approaches, the effects on the tail are
%  significantly larger than the scale variation bands of the LO calculation.''
%  There are no bands for the LO factorized calc beyond 150.}

 Measurements of $m_t$ are mostly affected by changes in shape around the peak
 of the distribution.
 For the full NLO calculation, there is a significant change in shape over the
 entire \mlb\ range (cf.~Figure~\ref{sfig:mlb2}). Moreover, this is
 the only prediction featuring a rather asymmetric uncertainty band.
 The upward and downward scale variations both lower the cross section
 in the range $80-150\gev$.
 This is in contrast to all LO calculations as well as to the NLO factorized
 calculation, for which the uncertainty bands are fairly symmetric around the
 respective central predictions.
 This behaviour is not caused by the choice of a dynamical scale,
 which has been verified by repeating the full calculations for the
 fixed scale $\mu=m_t$. In this case, a very similar behaviour was
 found.

 Since the shape variations are most relevant for the $m_t$ measurement,
 Figure~\ref{fig:mlbcompare} compares several normalized \mlb\ distributions.
 The effect of using different scales, namely $\mu=\hat H_T/2$, and $\mu=m_t$,
 is small. This is shown in Figure~\ref{sfig:mlbcompwwbb} for the full
 approach, and separately for the LO and NLO calculations.
 The evaluation of the $\hat H_T$ scale requires knowledge of the four-momenta
 of the top quark decay products. This is rather inconvenient when applied to
 factorized calculations.  
 Therefore, to directly compare the predictions of the full and
 factorized approach, the fixed scale of $\mu=m_t$ is used in
 Figure~\ref{sfig:mlbcomptt}.
 Apart from the above discussed differences for $\mlb>150\gev$, it is the
 shape of the full NLO prediction that deviates considerably by up to $20\%$
 from all other predictions.
 Given that this difference occurs in a region of large cross section, it will
 have a visible consequence for the top quark mass measurement discussed below.

 Once NLO corrections to the top quark pair production are incorporated,
 the \mlb\ distribution develops a tail. Already the non-resonant contributions
 included in the LO full calculation lead to a more pronounced tail, which in
 addition receives large NLO corrections.
 Even though the tail of the \mlb\ distribution plays only a minor role in the
 top quark mass determination, it is important to assess its impact, especially
 when aiming at a top quark mass measurement with a precision below $1\gev$.

\boldmath
\subsubsection{Investigation of theoretical uncertainties in the $\mt$
  measurement}
%\subsubsection{Top quark mass extraction based on pseudo-experiments}
\unboldmath

 The top quark mass measurement presented in Ref.~\cite{ATLAS-CONF-2013-077}
 uses a template method. For details of the implementation, see
 Ref.~\cite{AMaierDiplom:2012}.
 In short, in this method, simulated distributions are constructed for different
 input values of the top quark mass, \mtin.
 The distributions (templates) per \mtin\ are then individually fitted to a
 function. Using templates at different \mtin, it is verified that all
 parameters of the function linearly depend on $\mt=\mtin$. Consequently, this
 linearity is imposed in a combined fit to all templates. This fit fixes the
 theory model (i.e.~the parametrization of the theory model or hypothesis) by
 determining all parameters of the function, except for \mt, which is to be
 determined from data.
 Using those parameter values, a likelihood fit of this function to data is
 performed to obtain the value for \mt\ that best describes the data,
 namely \mtou, together with its statistical uncertainty. Using different sets
 of pseudo-data, the same strategy is used to estimate the impact of different
 theory descriptions. The systematic uncertainties on \mt\ stemming from
 theoretical uncertainties are mostly obtained by changing parameters in the
 Monte Carlo simulation, and assessing the shift of the fitted value of \mtou\
 while keeping the original template fit function.

%------------------------------------------------------------------
\begin{figure}[t!]
  \centering
  \includegraphics[width=0.7\textwidth]{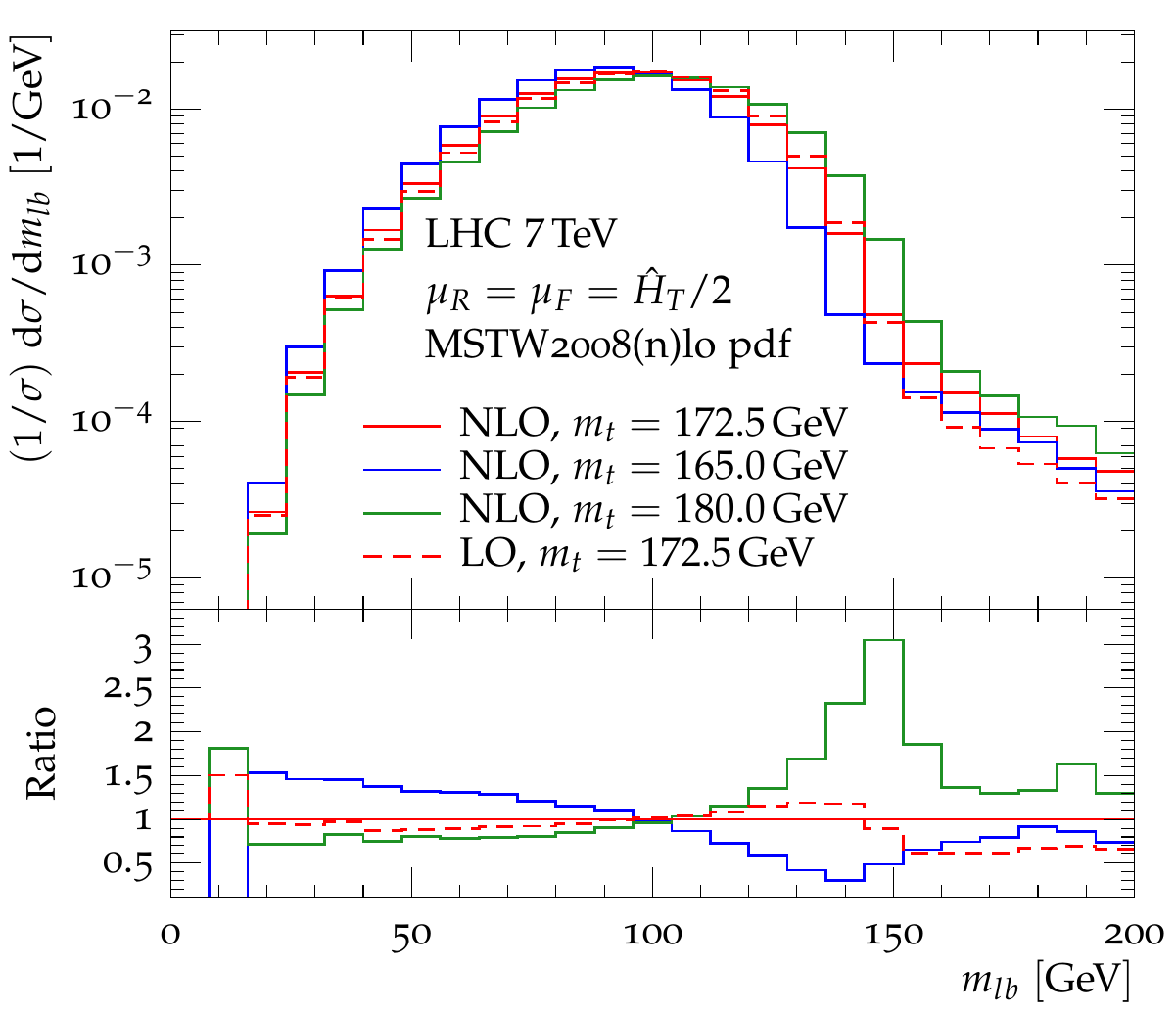}
  \caption{\label{fig:mlb}
    The normalized parton level \mlb\ distribution calculated at NLO in the full
    approach for three different top quark masses, utilizing the $\mu=\hat
    H_T/2$ scale. Also shown is a comparison to the LO prediction
    (dashed line) obtained at the default top quark mass value, which
    is $m_t=172.5\gev$.}
\end{figure} 
%------------------------------------------------------------------

 In experimental analyses, these templates are constructed at the detector
 level, i.e.~mimicking real data. Here, an analogous procedure is employed to
 assess the impact of the NLO corrections and their theoretical uncertainties on
 the \mlb\ method used to determine the top quark mass. We follow as closely as
 possible the procedure detailed in Ref.~\cite{ATLAS-CONF-2013-077}. More
 specifically, the goal of this study is to identify the size of mass shifts,
 which one can expect solely from scale variations of the NLO theory.
 Therefore, the templates are constructed and analyzed at the parton
 level. This means that any smearing occurring from the parton level
 to the detector level, i.e.~from the simulated to the observed
 distributions, cannot be addressed here.
 However, in this procedure, the top quark mass, $m_t$, can be
 identified with the top quark pole mass, since in our investigations,
 we only rely on parton level perturbative calculations.
 To illustrate the sensitivity of \mlb\ to the top quark mass, we show the
 normalized \mlb\ distributions for three different values of \mt\ in
 Figure~\ref{fig:mlb}. 
 Comparing the observed differences to Figure~\ref{sfig:mlbcomptt}
 reveals why shape changes of the order of $20\%$ will significantly
 influence the measurement of \mt.

\begin{figure}[t!]
  \centering
  \begin{subfigure}[b]{0.49\textwidth}
    \centering
    \includegraphics[width=1\textwidth]{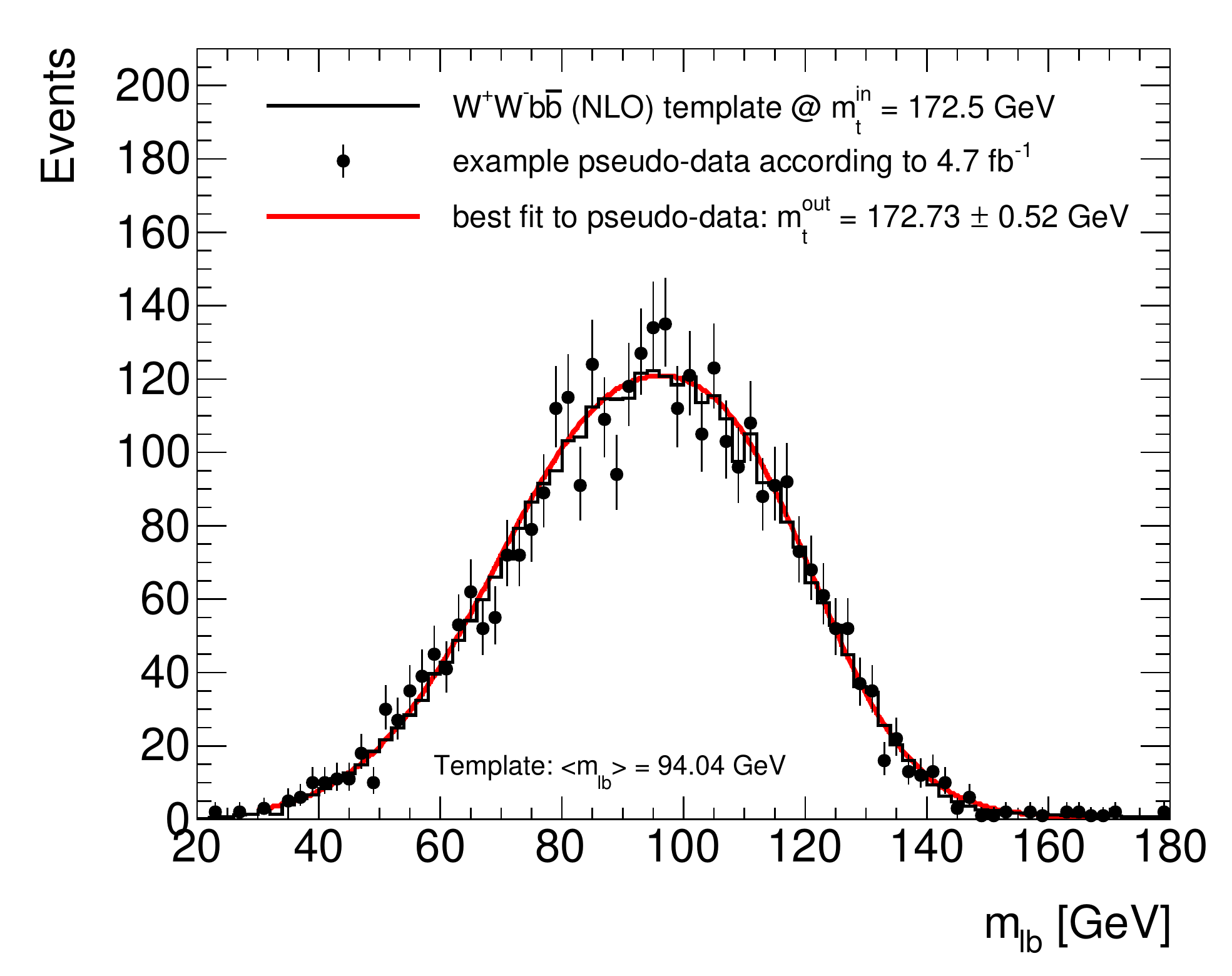}
    \caption{\label{sfig:atlasfit}}
  \end{subfigure}
  \hskip1mm
  \begin{subfigure}[b]{0.49\textwidth}
    \centering
    \includegraphics[width=1\textwidth]{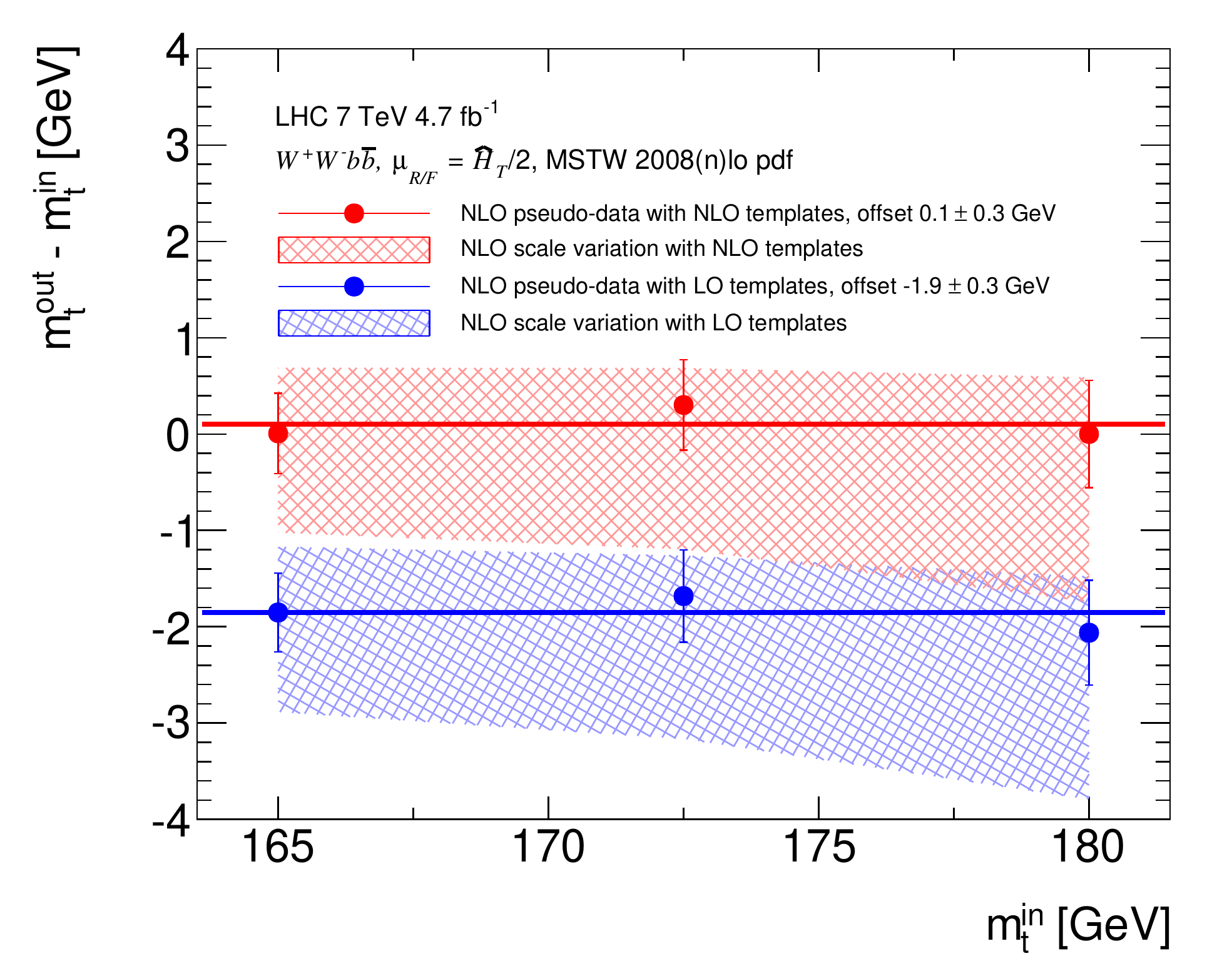}
    \caption{\label{sfig:atlasoffset}}
  \end{subfigure}
  \caption{\label{fig:mlbAtlas}
 Results from pseudo-data sets generated from the NLO calculation in the full
 approach. In (\ref{sfig:atlasfit}), one NLO pseudo-data set~(black points) at
 $\mtin=172.5\gev$ is shown together with its fit~(red line) and the underlying
 NLO template~(black histogram). The mean value of \mlb\ for the template is
 denoted by $\langle\mlb\rangle$.
 The predictions regarding the difference $\mtou-\mtin$ (i.e.~the \mt\
 offset) are depicted in (\ref{sfig:atlasoffset}) for three input
 values of \mt. These results were computed from many pseudo-data sets
 analyzed with a theory model based on NLO (red) or LO (blue) templates
 constructed from full calculations, and using $\mu=\hat H_T/2$.
 The points show the observed mean differences in $\mtou-\mtin$, together with
 their statistical uncertainty corresponding to a luminosity of
 $4.7/\mrm{fb}$. The horizontal lines stem from a fit of the three points to a
 constant, displaying the average offset.
 The bands indicate the offset observed when replacing the NLO pseudo-data by
 the ones obtained from the NLO scale variation samples.}
\end{figure}

\begin{figure}[t!]
  \centering
  \begin{subfigure}[b]{0.49\textwidth}
    \centering
    \includegraphics[width=1\textwidth]{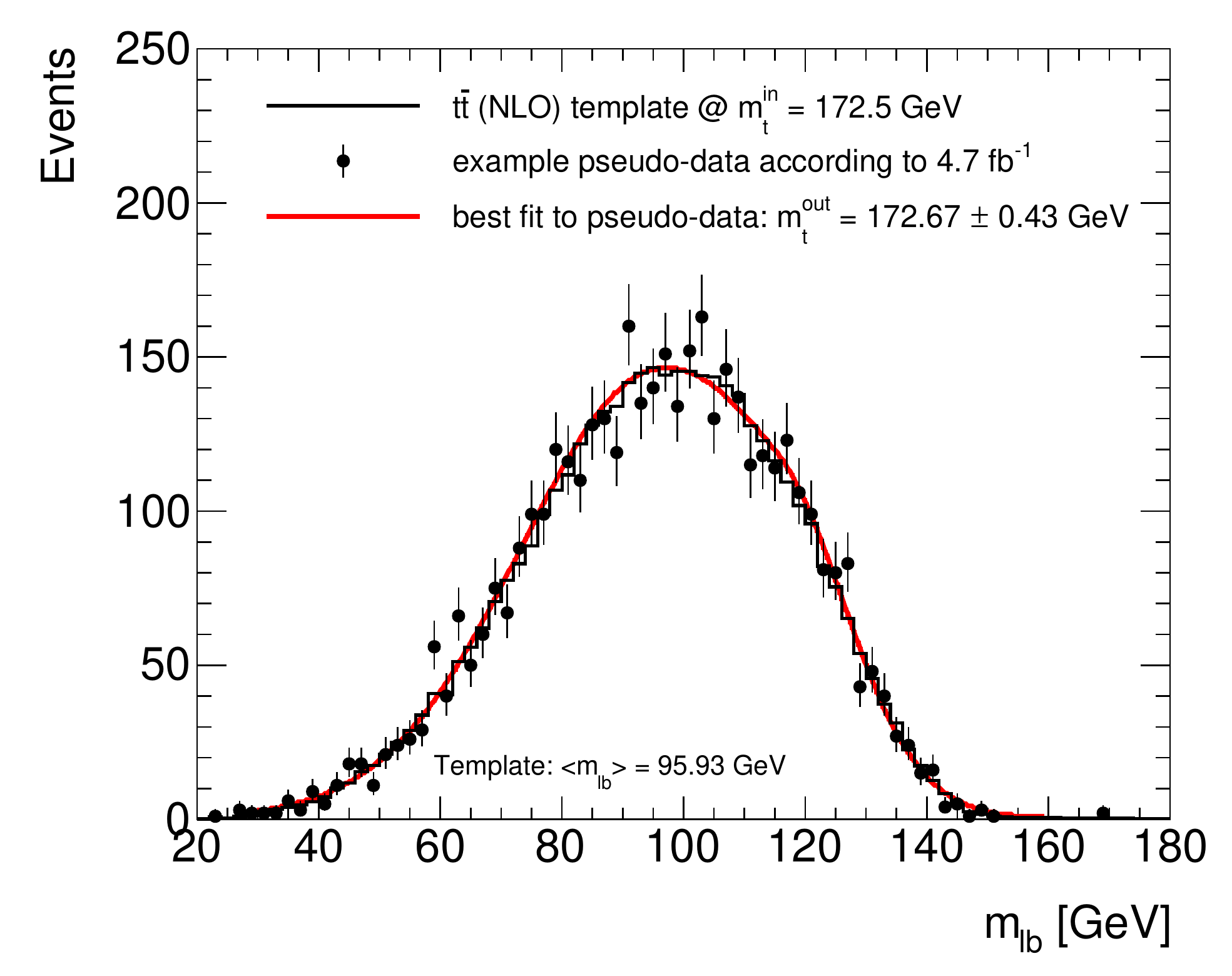}
    \caption{\label{sfig:atlasfittt}}
  \end{subfigure}
  \hskip1mm
  \begin{subfigure}[b]{0.49\textwidth}
    \centering
    \includegraphics[width=1\textwidth]{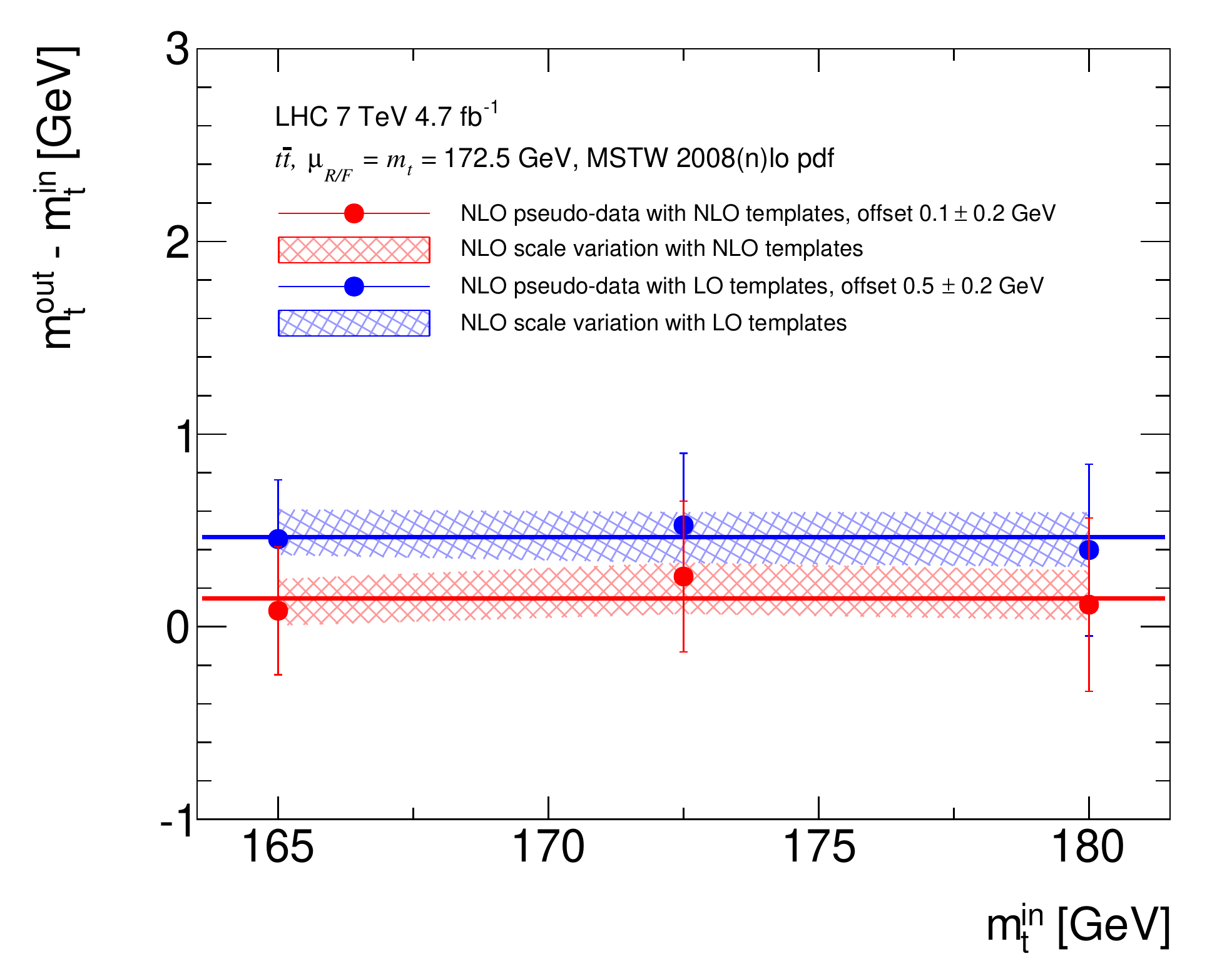}
    \caption{\label{sfig:atlasoffsettt}}
  \end{subfigure}
  \caption{\label{fig:mlbAtlastt}
    Same as Figure~\ref{fig:mlbAtlas}, but for pseudo-data sets and
    templates generated from the factorized calculations using $\mu=m_t$.
    Note that the vertical axes' ranges of both figures are different
    from the corresponding ones in Figure~\ref{fig:mlbAtlas}.}
\end{figure}

 In this analysis, the pseudo-data mimicking experimental data (i.e.~the data
 model) are always generated from the NLO predictions simulating a data
 luminosity of $4.7/\mrm{fb}$ as was analyzed in
 Ref.~\cite{ATLAS-CONF-2013-077}.
 In contrast, the templates (i.e.~the theory hypothesis) are either taken from
 the NLO or the LO predictions (using the same scale settings), and are referred
 to as NLO or LO templates, respectively.
 Because we have found sizeable differences in the predicted shape of
 the \mlb\ distribution between the full and factorized approach, we have
 performed our investigations separately, relying on either the full
 calculations or the factorized ones.
 For each of these calculational scenarios, we investigate the impact on the top
 quark mass measurement caused by two aspects, namely the scale variations and
 the shape modifications arising from NLO corrections.
%\footnote{instead of ``predicted LO to NLO differences.''}
 To study the latter aspect, we switch from the NLO to the LO
 description of our template theory model.
 The results for the full calculations using $\mu=\hat H_T/2$ are summarized in
 Figure~\ref{fig:mlbAtlas}, while Figure~\ref{fig:mlbAtlastt} displays those of
 the factorized approach at $\mu=m_t$.
 The two scenarios are discussed in turn.

%------------------------------------------------------------------
%------------------------------------------------------------------

 The points in Figure~\ref{sfig:atlasfit} show a pseudo-data set, i.e.~one
 possible experimental outcome. This pseudo-data set was generated from the NLO
 prediction of Figure~\ref{fig:mlb} at $\mtin=172.5\gev$, which in this figure
 is shown as the (black) histogram.
 The result of the template fit using the NLO templates is displayed as the
 (red) line in this figure.
 Within the sizeable variations of the data points given the data statistics,
 the fit coincides with the underlying theory hypothesis, demonstrating the
 internal consistency of the method. Given these uncertainties, with the
 presently available luminosity, the shape differences of the LO and NLO
 templates seen in Figure~\ref{fig:mlb} cannot be discriminated from
 experimental data, but will be compensated for by a different fit-value
 obtained for \mt.

 The sensitivity to the theoretical assumptions and their uncertainties is
 assessed by fits to one thousand pseudo-data sets.
 For three different values of \mtin, Figure~\ref{sfig:atlasoffset} shows the
 observed difference of \mtou, the mass measured by the procedure, and \mtin,
 the one used to generate the pseudo-data.
 The red points correspond to the mean difference observed for all pseudo-data
 sets that are produced as in Figure~\ref{sfig:atlasfit} and analyzed with the
 NLO templates.
%As expected, all points are compatible with the assumption of a
%vanishing \mt\ offset.
 The uncertainty per point is statistical only and corresponds to the
 expected experimental uncertainty for the assumed data luminosity.
 The red band corresponds to the scale uncertainty on the measured top quark
 mass obtained by replacing the pseudo-data with those from the scale variation
 NLO samples, while keeping the original NLO templates. 
 The resulting uncertainty is significantly larger than the statistical
 precision, and of similar size as the total theoretical systematic uncertainty
 assigned to the experimental result~\cite{ATLAS-CONF-2013-077}.

 It has been shown in Figures~\ref{sfig:mlb2} and~\ref{fig:mlb} that, at the
 same top quark mass, the predicted \mlb\ distributions calculated in the full
 approach at LO and NLO are significantly different.
 The flatness of the leading order scale variation band in
 Figure~\ref{sfig:mlb2} shows that the LO scale variations -- although strongly
 affecting the cross section -- introduce only small shape distortions into
 the \mlb\ distribution.
 Given that the shape changes observed at NLO are significant, determining shape
 dependent observables assuming LO predictions as theory model will inevitably
 suffer from this shortcoming of the LO prediction.
 Nevertheless, to assess the size of the effect for full calculations, we also
 performed a determination of the top quark mass using the LO templates, still
 based on the NLO pseudo-data sets.
 This mimics the situation in which LO templates are used to measure the top
 quark mass from data that actually resemble the NLO prediction.
 The result of this is shown as blue points in Figure~\ref{sfig:atlasoffset}.
 In this parton level investigation, the difference $\mtou-\mtin$
 turns out to be about $-1.9\gev$. Consequently, a sizeable (but
 different) offset is expected when using LO predictions for
 experimental top quark mass measurements.
 In this situation, the data would also suffer from the scale variation
 uncertainties, as can be seen from the blue band, obtained by generating the
 pseudo-data sets in the same way as for the red band, but keeping the LO
 templates as theory model.
 Clearly, although not properly assessable within the LO description of the 
 theory model, the effect would be present in data.

%------------------------------------------------------------------
%------------------------------------------------------------------

 Using exactly the same strategy, the results for the factorized
 approach are shown in Figure~\ref{fig:mlbAtlastt}.
 Again, the implications of scale variations and template modelling at
 different orders in $\alpha_\mrm{s}$ are discussed in turn.
 For the factorized calculation, the size of the red band reflecting the scale
 variation uncertainty is about $\pm0.2\gev$, which is significantly smaller
 than the one of the full calculation, where it amounts to about
 $^{+0.6}_{-1.0}\gev$. This is a direct consequence of the distinct
 sizes of the observed shape differences in Figures~\ref{fig:mlbnlo}
 and \ref{fig:mlbcompare}.
 The \mlb\ shape variations predicted by the factorized calculations
 at NLO are very small.
%As noted, the \mlb\ shape as predicted by the factorized NLO
%calculations is fairly stable.
%
 However, this only happens because important effects from non-factorizing
 contributions and higher-order corrections to the top quark decays are not
 captured by this approximation.
 Consequently, the larger uncertainty on \mt\ observed for the full calculation
 is certainly more realistic than the small one predicted when using the
 factorized approach.
 The mean values $\langle\mlb\rangle$ of the NLO templates shown in
 Figures~\ref{sfig:atlasfit} and \ref{sfig:atlasfittt}, which are
 obtained for the same \mtin, are different by almost $2\gev$.
 This is a manifestation of the sizeable differences in the predicted \mlb\
 distributions for the two NLO calculations. In addition, the predicted cross
 section of the factorized approach is about $20\%$ higher than the
 one from the full approach.

 To investigate the differences between the NLO and LO description in
 the factorized approach, we follow the same procedure as for the full
 calculation, and use NLO pseudo-data together with LO templates.
 The results are presented in Figure~\ref{sfig:atlasoffsettt}, and as
 before, they are shown in blue.
 For the factorized approach, we observe a much smaller scale variation band than
 for the full approach shown in Figure~\ref{sfig:atlasoffset}.
 In addition, a much reduced \mt\ offset is found from analyzing NLO
 pseudo-data with LO templates. It amounts to only about $0.5\gev$.
 Both effects are caused by the behaviour of the \mlb\ distributions
 presented in Figures~\ref{sfig:mlbnwa} and \ref{sfig:mlbcomptt}. In
 the region of largest sensitivity, the NLO corrections alter
 the \mlb\ shape as given at LO to a much smaller extent than what is
 observed for the full calculation.
 For the factorized approach, sizeable shape differences only occur in the high
 mass tail of the \mlb\ distribution, whose impact is largely suppressed
 compared to the peak region due to the small cross section.

%\subsection{Asymmetries}%\label{sec:asym}
\subsection{Top quark asymmetries}\label{sec:asym}

Having both the full and factorized NLO computation
(cf.~Section~\ref{sec:toptreat}, item (I) and (II), respectively) for
the production of the $W^+W^-b\bar b$ final state at hand, we are in a
convenient position to take a closer look at how finite width effects
and non-factorizing contributions impact top quark asymmetries as
measured at the LHC and at the Tevatron~\cite{Aaltonen:2011kc,Abazov:2011rq,CDF:2011vba,Aaltonen:2012it,CDF:2013gna,Aaltonen:2013vaf,Abazov:2013wxa,ATLAS:2012an,ATLAS-CONF-2012-057,Aad:2013cea,Chatrchyan:2011hk,Chatrchyan:2012cxa,CMS-PAS-TOP-12-010,CMS-PAS-TOP-12-033}.
We treat the leptonic decays of the $W$ bosons in a way such that the
spin correlations are preserved. This allows us to study to what
extent the lepton-based asymmetries inherit the effects on top quark
asymmetries.

The symmetry of the LHC's $pp$ initial state makes it impossible to
write down a forward-backward asymmetry variable as known from the
Tevatron experiments. The difference in the broadness of the
respective rapidity distributions for the top quark ($y_t$) and
the antitop quark ($y_{\bar t}$) can however be exploited to define
the (commonly used) charge asymmetry:
\begin{equation}
  \label{eq:cast}
  A^\srm{C}_{t\bar t}\;=\;
  \frac{\sigma\left(\upDelta\left|y\right|>0\right) -
        \sigma\left(\upDelta\left|y\right|<0\right)}
       {\sigma\left(\upDelta\left|y\right|>0\right) +
        \sigma\left(\upDelta\left|y\right|<0\right)}~.
\end{equation}
Using $\upDelta\left|y\right|=\left|y_t\right|-\left|y_{\bar t}\right|$,
this asymmetry is accessible to the LHC experiments. In fact it
has been measured by the ATLAS~\cite{ATLAS:2012an,Aad:2013cea} and
CMS~\cite{Chatrchyan:2011hk,Chatrchyan:2012cxa,CMS-PAS-TOP-12-033}
collaborations. Based on the current results, the asymmetry is found
to be in agreement with the Standard Model predictions. The leptonic
charge asymmetry in the dilepton channel, $A^\srm{C}_{ll}$, is defined
analogously, where one replaces $\upDelta\left|y\right|$ in
Eq.~\eqref{eq:cast} by
$\upDelta\left|\eta\right|=\left|\eta_{l^+}\right|-\left|\eta_{l^-}\right|$.

%\begin{equation}
%A_{l^+l^-}^{C}=\frac{\sigma\left(\Delta \left|\eta \right|>0\right) -  \sigma\left(\Delta \left|\eta \right|<0\right)}{ \sigma\left(\Delta \left|\eta \right|>0\right) + \sigma\left(\Delta \left|\eta \right|<0\right)}
%\label{eq:casl}
%\end{equation}
%with $\Delta \left|\eta \right| = \left|\eta_{l^+} \right| - \left|\eta_{l^-} \right|$.
%For a recent review on the status of the top charge asymmetry we
%refer to \cite{Westhoff:2013ixa}.

Using the kinematic constraints detailed in Section~\ref{sec:res}, we
summarize our predictions based on the full approach for the top quark
and leptonic charge asymmetry below, reading
\be
A^\srm{C}_{t\bar t}\;=\;0.008\pm0.003\quad\text{and}\quad
A^\srm{C}_{ll}\;=\;0.005\pm0.003\,,
\ee
respectively. The theoretical uncertainties of these asymmetry values
have been estimated from standard scale variations. We found good
agreement with the results stated in Ref.~\cite{Denner:2012yc}.
%To indicate the magnitude of
%variations possible at NLO, we also show values obtained for different
%scale choices and in the factorized approach.
The currently measured experimental values in the dilepton channel are
$A^\srm{C}_{t\bar t}=0.057\pm0.028$ and
$A^\srm{C}_{ll}=0.023\pm0.014$ (ATLAS, $7\tev$)~\cite{ATLAS-CONF-2012-057}
as well as $A^\srm{C}_{t\bar t}=0.050\pm0.043^{+0.010}_{-0.039}$ and
$A^\srm{C}_{ll}=0.010\pm0.016$ (CMS, $7\tev$)~\cite{CMS-PAS-TOP-12-010}.
%For completeness, we have added the current experimental results from
%ATLAS and CMS.
Note that the comparison to these values can only be of qualitative
nature, owing to the different kinematical selections used
in our study as opposed to the experiments.

The relation between top quark and leptonic asymmetries has also been
studied at the Tevatron where, owing to the $p\bar p$ initial state,
forward-backward quantities are very meaningful~\cite{Aaltonen:2011kc,Abazov:2011rq,Aaltonen:2012it,Aaltonen:2013vaf,Abazov:2013wxa}.
In fact, these variables are more sensitive to the underlying
asymmetry effect.
%The impact of these measurements however is limited as a result of the lower
%statistics that was obtained at the Tevatron.
The top quark forward-backward asymmetry $A^\srm{FB}_{t\bar t}$ is
defined as
\begin{equation}
  \label{eq:fbast}
  A^\srm{FB}_{t\bar{t}}\;=\;
  \frac{\sigma\left(\upDelta y>0\right)-\sigma\left(\upDelta y<0\right)}
       {\sigma\left(\upDelta y>0\right)+\sigma\left(\upDelta y<0\right)}
\end{equation}
where $\upDelta y=y_t-y_{\bar t}$ and $y_t$ again denotes the rapidity
of the top quark. For $t\bar t$ production calculated at leading order
in QCD, this asymmetry vanishes. The first non-zero contribution to
$A^\srm{FB}_{t\bar t}$ appears at NLO. 
%but is too small to be in agreement with the experimental findings. 
Measurements of $A^\srm{FB}_{t\bar t}$ at the Tevatron~\cite{Aaltonen:2011kc,Abazov:2011rq,CDF:2011vba,Aaltonen:2012it,CDF:2013gna}
give significantly larger values than the Standard Model prediction
\cite{Kuhn:2011ri,Hollik:2011ps,Bernreuther:2012sx,Kuhn:2013zoa}. 
To help resolve the discrepancy, several suggestions have been made
with the aim to obtain additional handles in the measurements, cf.~for
example Refs.~\cite{AguilarSaavedra:2012va,Skands:2012mm,Falkowski:2012cu,Winter:2013xka,Hoeche:2013mua,Berge:2013xsa,Li:2013mia,Gripaios:2013rda}.

The disadvantage of the top quark asymmetry is that it cannot be
measured directly. As the top quarks have to be reconstructed from
their decay products and the missing transverse momentum, all
experimental results on $A^\srm{FB}_{t\bar t}$ depend on the respective
reconstruction method. To avoid any potential bias introduced as a
result of the kinematic procedure employed to reconstruct the top
quark momenta, several lepton-based asymmetries were designed which
only depend on the object selection. The drawbacks of the leptonic
asymmetries are their decay-channel specific definition and lower
sensitivity to detect the actual asymmetry.
%one obtains in using them. 
The effect as seen in $A^\srm{FB}_{t\bar t}$ will be washed
out. However, as for example pointed out in
Ref.~\cite{Falkowski:2012cu}, this becomes less of an issue once the
top quarks are sufficiently boosted such that the leptons mostly 
follow the respective top quark directions. As shown in
\cite{Falkowski:2012cu}, a convenient quantity to tune this
correlation between $A^\srm{FB}_{t\bar t}$ and lepton-based
asymmetries is given by the lepton transverse momentum, $p_{T,l}$.

%To calculate the leptonic forward-backward asymmetry, an additional run was made for $p\bar{p}$ collisions at $\sqrt{s}=1.96$\,TeV Tevatron. The cuts are slightly different from the LHC ones, in that the jet separation is $\Delta R=0.4$, the $b$ jets only have to have $p_{T,b}>20$\,GeV, and the cut for the missing momentum is $p_{T,\text{miss}}>20$\,GeV. The scale $\mu$ is chosen as
%\begin{equation}
%\mu=\mu_R=\mu_F=m_t.
%\label{eq:fixedscaletev}
%\end{equation} 
%This yields a leptonic forward-backward asymmetry of 
%(for the definition see section \ref{sec:asym}) of 
%\begin{equation}
%A_{l^+l^-}^{FB}=0.036\pm24\% \text{(stat.)}
%\label{eq:fbaresult}
%\end{equation} 
%which is in good agreement with the result of \cite{Denner:2012yc} $A_{l^+l^-}^{FB}=0.0361(5)$. 
%A more detailed study of differential forward-backward asymmetries is given in Section \ref{sec:asym}.
%Here we focus on distributions for observables where the NLO corrections are particularly 
%important as they change the shape of the distributions. 

%It is also instructive to look at differential asymmetries. 
%The differential leptonic forward-backward asymmetry is defined as
%\begin{equation}
%A_{l^+l^-}^{FB}(O)=\frac{\frac{\text{d}\sigma}{\text{d}O}\left(\Delta \eta>0\right) -  \frac{\text{d}\sigma}{\text{d}O}\left(\Delta \eta<0\right)}{ \frac{\text{d}\sigma}{\text{d}O}\left(\Delta \eta>0\right) + \frac{\text{d}\sigma}{\text{d}O}\left(\Delta \eta<0\right)}.
%\label{eq:fbdiff}
%\end{equation}
%By choosing different observables for $O$, one can study the
%dependence of the asymmetry on these.

\begin{figure}[t!]
  \centering
  \begin{subfigure}[b]{0.3\textwidth}
    \centering
    \includegraphics[width=1\textwidth]{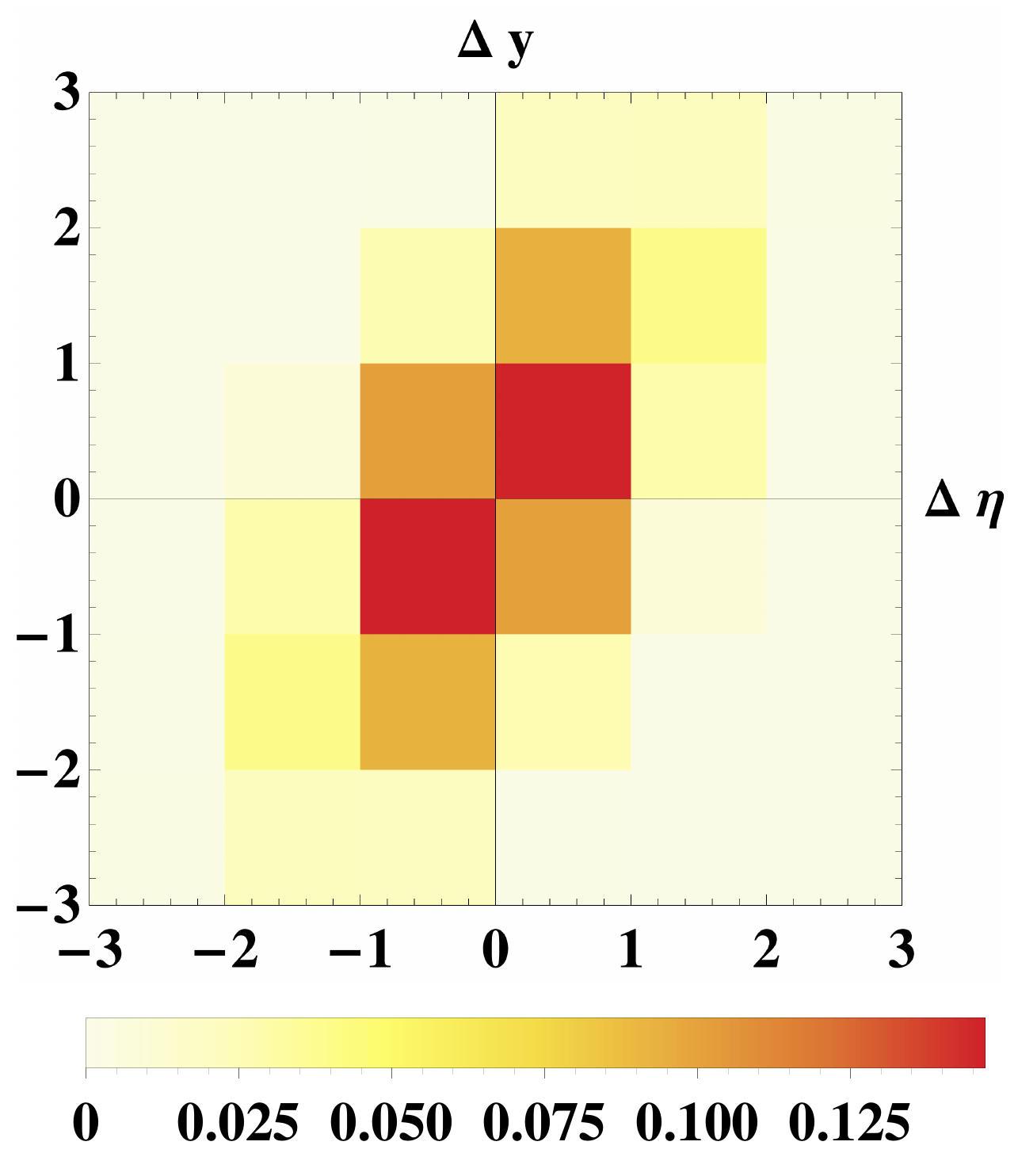}
    \caption{\label{sfig:lo2d} LO}
  \end{subfigure}
  \hskip5mm
  \begin{subfigure}[b]{0.3\textwidth}
    \centering
    \includegraphics[width=1\textwidth]{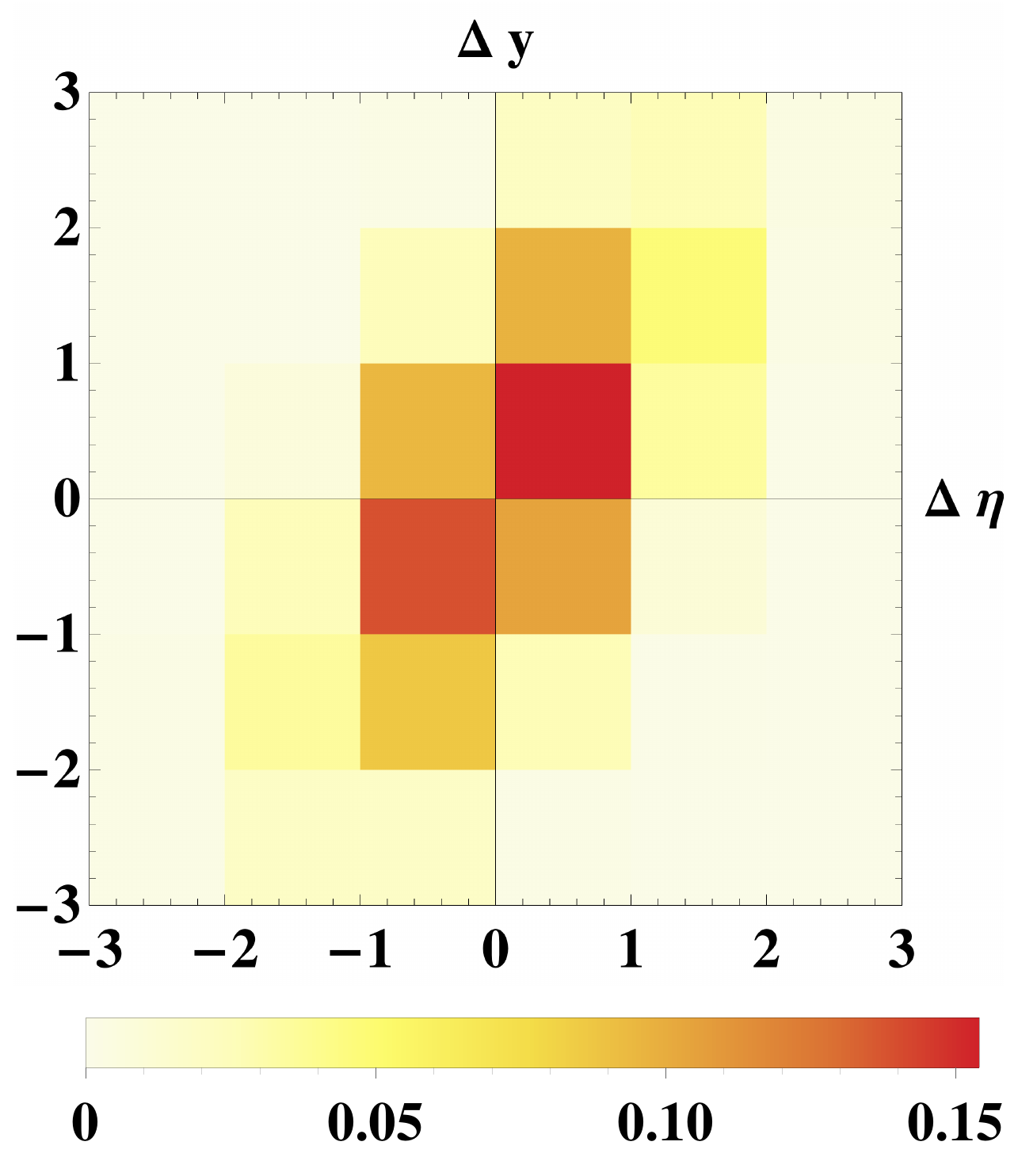}
    \caption{\label{sfig:nlo2d} NLO}
  \end{subfigure}
  \hskip5mm
  \begin{subfigure}[b]{0.3\textwidth}
    \centering
    \includegraphics[width=1\textwidth]{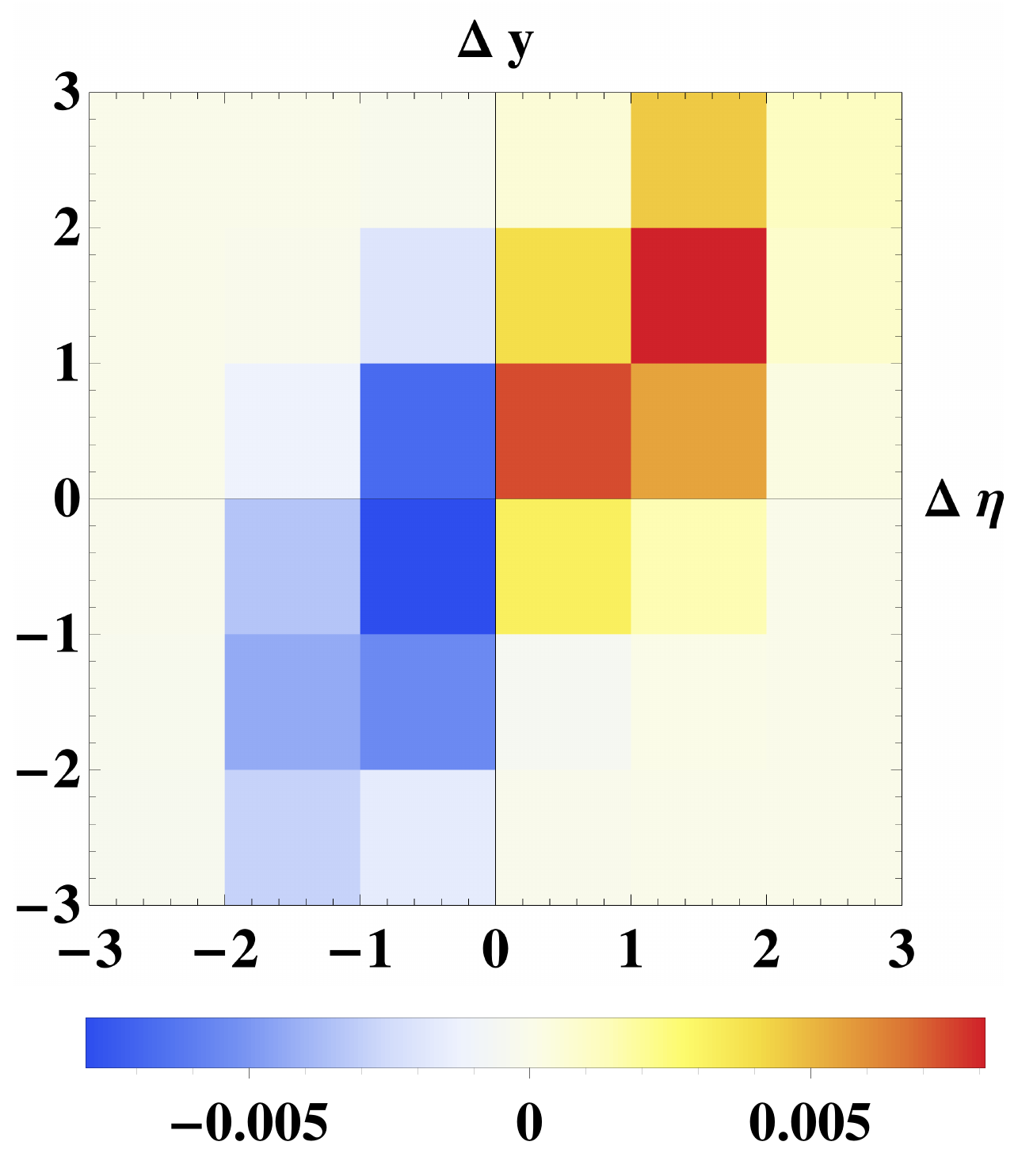}
    \caption{\label{sfig:diff2d} NLO minus LO}
  \end{subfigure}
  \caption{\label{fig:3d}
    Double differential cross section
    $\frac{1}{\sigma}\frac{d\sigma}{d\upDelta\eta\,d\upDelta y}$ at
    the Tevatron in dependence on the rapidity differences
    $\upDelta y$ and $\upDelta\eta$. Dilepton channel predictions for
    $W^+W^-b\bar b$ production in the full approach normalized by the
    respective total cross section are shown at LO (\ref{sfig:lo2d})
    and NLO (\ref{sfig:nlo2d}). The difference between the individual
    NLO and LO two-dimensional shapes is also visualized
    (\ref{sfig:diff2d}).}
\end{figure}

Our calculations concern the dilepton channel, for which we can employ
a commonly used leptonic asymmetry that reads
\begin{equation}
  \label{eq:fbasl}
  A^\srm{FB}_{ll}\;=\;
  \frac{\sigma\left(\upDelta\eta>0\right)-\sigma\left(\upDelta\eta<0\right)}
       {\sigma\left(\upDelta\eta>0\right)+\sigma\left(\upDelta\eta<0\right)}~.
\end{equation}
It is based on defining $\Delta\eta=\eta_{l^+}-\eta_{l^-}$ where
$\eta_{l^\pm}$ denotes the pseudo-rapidity of the charged leptons. To
get a better understanding of the relation between $A^\srm{FB}_{t\bar t}$
and $A^\srm{FB}_{ll}$ in $p\bar p$ collisions at $\sqrt{s}=1.96\tev$,
we first consider the normalized double differential cross section
depending on both rapidity difference measures, i.e.\
$\frac{1}{\sigma}\frac{d\sigma}{d\upDelta\eta\,d\upDelta y}$. We use
the fixed scale choice $\mu=m_t$, with $m_t=172.0\gev$, and a
$b$-flavour sensitive anti-$k_T$ jet algorithm that allows us to keep
track of the total $b$ charge contained in the jet, as suggested in
Refs.~\cite{Banfi:2006hf,Banfi:2007gu}, and impose kinematical
requirements reading:
\be
\label{eq:ewparam2}
\begin{matrix}
  \upDelta R &>& 0.4\,,~~~~~\quad & |\eta_b| &<& 2.5\,,~~~~~\quad &
  |\eta_l| &<& 2.5\,,~~~~~\\
  \slashed{p}_T &>& 25\gev\,,\quad     & p_{T,b}   &>& 20\gev\,,\quad &
  p_{T,l}   &>& 20\gev\,.
\end{matrix}
\ee
To reconstruct the top quarks in our parton level simulation, we
first recombine the four final state leptons into the two $W$ bosons
according to the Monte Carlo information. We then employ the $b$ jet
charge information, which we get from the $b$-flavour specific jet
algorithm, to assign exactly one $b$ jet to each $W$ boson. Events
that cannot be analyzed this way are disregarded. Additional jets
arising from real radiation are tested kinematically whether they
belong to the pseudo-top or pseudo-antitop quark
candidate.\footnote{Other procedures neglecting $b$-jet truth
  information were tested, including one based on the sole kinematic
  reconstruction of the top quark objects. In this parton level
  analysis, we however did not observe any significant changes in our
  results.}
The results which we obtained this way at LO and NLO in the full
approach are shown in Figure~\ref{fig:3d}, which is also used to
visualize the relative difference between these normalized
distributions (cf.~Figure~\ref{sfig:diff2d}). Based on
Figure~\ref{fig:3d}, the impact of the NLO corrections on $\upDelta y$
and $\upDelta\eta$ can be assessed in a convenient manner. We observe
that the NLO corrections to the total cross section cause a shift
of both $\upDelta y$ and $\upDelta\eta$ to larger values, where
slightly stronger shifts are seen for $\upDelta y$.

\begin{figure}[t!]
  \centering
  \begin{subfigure}[b]{0.98\textwidth}
    \centering
    \includegraphics[width=1\textwidth]{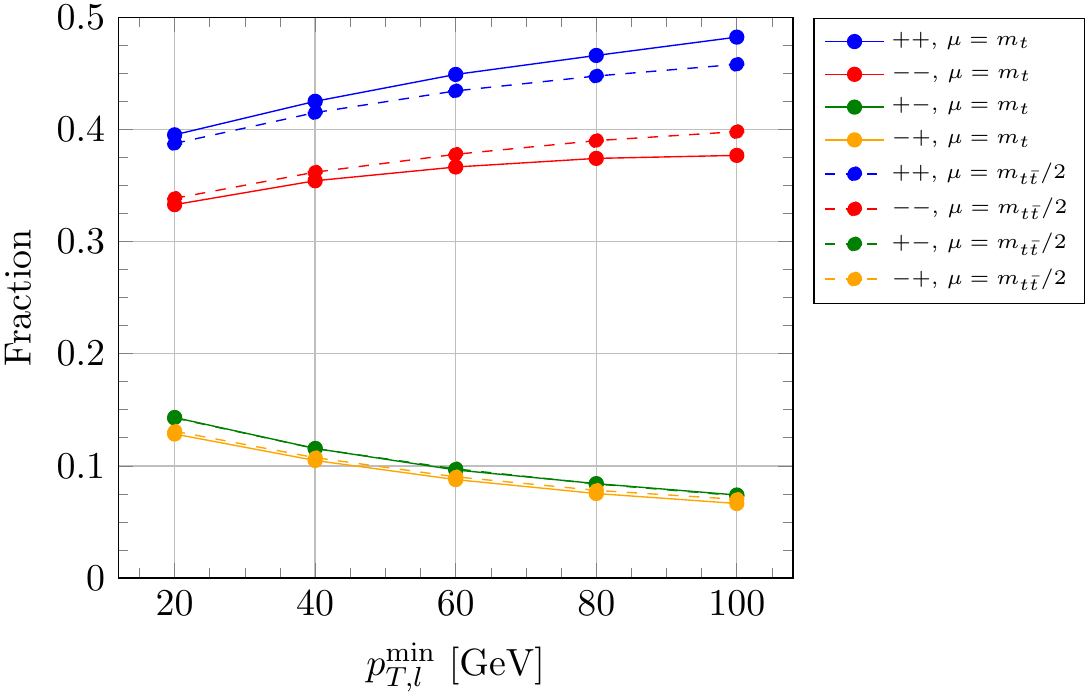}
  \end{subfigure}
  \caption{\label{fig:4field}
    Fractions of $W^+(e^+\nu_e)\,W^-(\mu^-\bar\nu_\mu)\,b\bar b$ events
    in four different kinematic areas defined by combinations of
    positive and negative $\upDelta y$ and $\upDelta\eta$ regions (see
    text for the details). Tevatron parton level predictions of NLO
    calculations in the full approach are shown for two different
    scale choices, $\mu=m_t$ and $\mu=m_{t\bar t}/2$.}
\end{figure}

Figure~\ref{fig:3d} emphasizes the importance of an accurate
description of final states containing $b$ jets, two oppositely
charged leptons and missing energy. Any statement based on how the
different asymmetries, $A^\srm{FB}_{t\bar t}$ and $A^\srm{FB}_{ll}$,
are correlated heavily relies on the robustness of the Standard Model
prediction for this final state, whose major contributor is
$W^+W^-b\bar b$ production including the leptonic decays. This
motivated us to investigate the behaviour of the
$A^\srm{FB}_{t\bar t}$ --- $A^\srm{FB}_{ll}$ correlation under
different minimal transverse momentum constraints,
$p^\srm{min}_{T,l}$, applied to the lepton momenta, thereby
re-visiting part of the ideas of Ref.~\cite{Falkowski:2012cu}.
Starting from %the idea as illustrated in
Figure~\ref{fig:3d}, we now simplify the parametrization of the
$\upDelta y$ --- $\upDelta\eta$ space by partitioning it into four
kinematic regions, which we label $++$, $+-$, $-+$ and $--$ according
to $\upDelta y>0~\&~\upDelta\eta>0$,
$\upDelta y>0~\&~\upDelta\eta<0$, $\upDelta y<0~\&~\upDelta\eta>0$ and
$\upDelta y<0~\&~\upDelta\eta<0$, respectively. The $\pm\mp$ bins may
suffer from the lower statistics, but the decomposition is simply a
compromise between experimental accessibility and theoretical detail
whose verification will be of great value for the experiments.
Figure~\ref{fig:4field} displays the fractions of (unweighted) events
populating these four bins for five different values of
$p^\srm{min}_{T,l}$. We show the results obtained for two different
scale choices in the full NLO approach. Both predictions -- the one
using the fixed scale $\mu=m_t$ and the one using the dynamical scale
$\mu=m_{t\bar t}/2$ where $m_{t\bar t}$ denotes the pair's invariant
mass -- give very similar results.\footnote{The evaluation of the
  $m_{t\bar t}$ scale proceeds via the identification of the two
  leading $b$ jets and combining them with the four leptons to the
  invariant mass of a $t\bar t$-like system.}
As expected, for increasing $p^\srm{min}_{T,l}$, the relative weight
of the $++$ and $--$ bins rises further.
% as compared to the $\pm\mp$ bins.

%This yields a leptonic forward-backward asymmetry of 
%\begin{equation}
%A_{l^+l^-}^{FB}=0.036\pm24\% \text{(stat.)}
%\label{eq:fbaresult}
%\end{equation} 

\begin{figure}[t!]
  \centering
  \begin{subfigure}[b]{0.98\textwidth}
    \centering
    \includegraphics[width=1\textwidth]{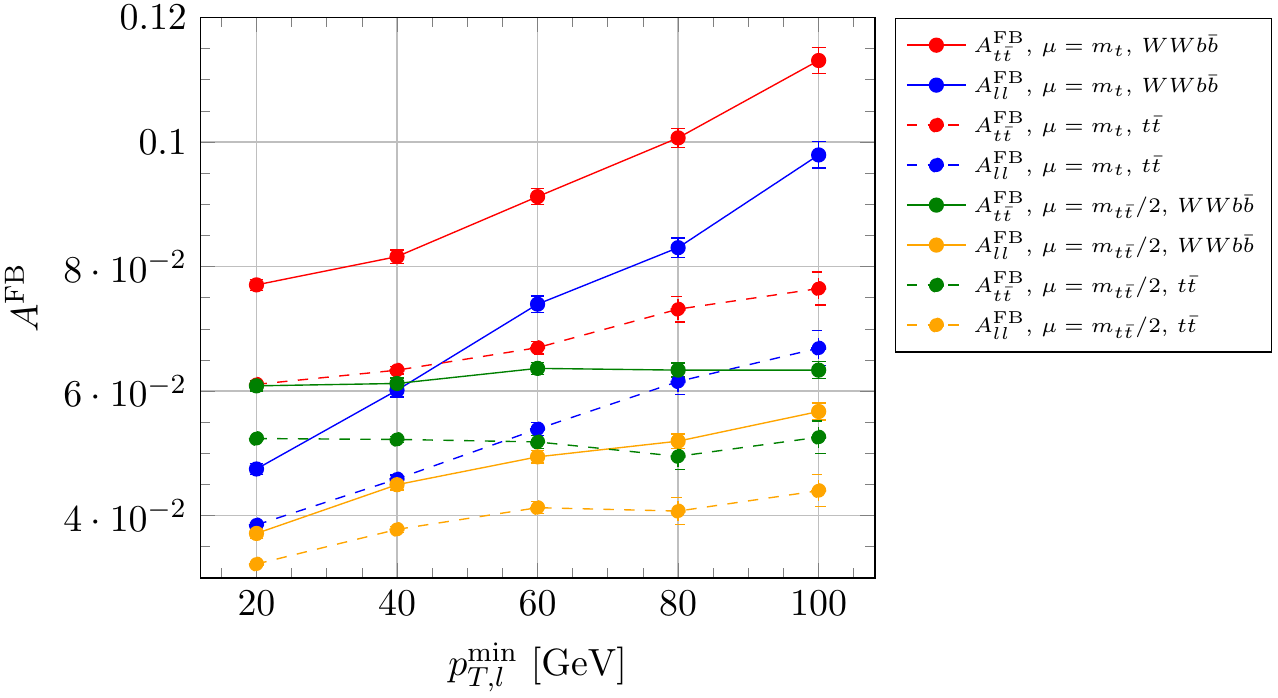}
  \end{subfigure}
  \caption{\label{fig:asymptmin}
    Dependence of the $t\bar t$ and leptonic forward-backward
    asymmetries on the kinematical requirement concerning the minimal
    charged lepton transverse momentum $p^\srm{min}_{T,l}$. Different
    lines belong to different scale choices; the solid and dashed
    lines respectively correspond to the full ($WWb\bar b$) and
    factorized ($t\bar t$) approach in calculating $W^+W^-b\bar b$
    production at NLO in the dilepton channel at the Tevatron. Both
    types of calculations are explained in Section~\ref{sec:toptreat}.
    The vertical bars denote the Monte Carlo statistical uncertainties.}
\end{figure}

To quantify this in terms of the asymmetries stated in
Eqs.~\eqref{eq:fbast} and \eqref{eq:fbasl}, we respectively evaluate
the dependence of $A^\srm{FB}_{t\bar t}$ and $A^\srm{FB}_{ll}$ on the
imposed minimal transverse momentum of the charged leptons. The
results are depicted in Figure~\ref{fig:asymptmin} for five
different values of $p^\srm{min}_{T,l}$. This time we have included
the predictions from the factorized approach to enable direct
comparison between the two calculational approaches, hence estimating
the effects missed by the factorized description. The scale choices
utilized to produce the results are, as before, $\mu=m_t$ and
$\mu=m_{t\bar t}/2$.
We observe, in accordance with the findings above (and within the
Monte Carlo statistics achieved) that the difference between
$A^\srm{FB}_{t\bar t}$ and $A^\srm{FB}_{ll}$ decreases with increasing
$p^\srm{min}_{T,l}$. It also becomes clear that the absolute change of
$A^\srm{FB}_{t\bar t}$ and $A^\srm{FB}_{ll}$ with $p^\srm{min}_{T,l}$
rather strongly depends on the scale choice. As can be seen,
increasing the $p_T$ threshold for the charged leptons causes the
asymmetries to rise faster once we rely on the fixed scale ($\mu=m_t$)
instead of the dynamical ones ($\mu=m_{t\bar t}/2$). Almost no rise
can only be found for the two top quark asymmetry predictions obtained
with the dynamical scale choice. These turn out to be rather constant
over the $p^\srm{min}_{T,l}$ range investigated here, but differ in
that the factorized prediction has dropped by $\sim15\%$ below the
full one. This trend is more general; in comparison to the NLO $t\bar t$
approach, the full $WWb\bar b$ treatment is found to generate
systematically larger asymmetries. Yet, the difference is not
sufficient to reconcile the theory predictions with the current
experimental measurements. Taking the result for the lowest
$p^\srm{min}_{T,l}$ cut and the $m_t$ scale choice, we note good
agreement with the results stated in Ref.~\cite{Bernreuther:2012sx},
although slightly tighter pseudo-rapidity constraints (namely
$|\eta_{b,l}|\le2.0$, cf.~Eqs.~\eqref{eq:ewparam2}) were used in this
work. For the experimental status, see
Refs.~\cite{CDF:2011vba,Aaltonen:2013vaf,Abazov:2013wxa}.

\begin{figure}[t!]
  \centering
  \begin{subfigure}[b]{0.98\textwidth}
    \centering
    \includegraphics[width=1\textwidth]{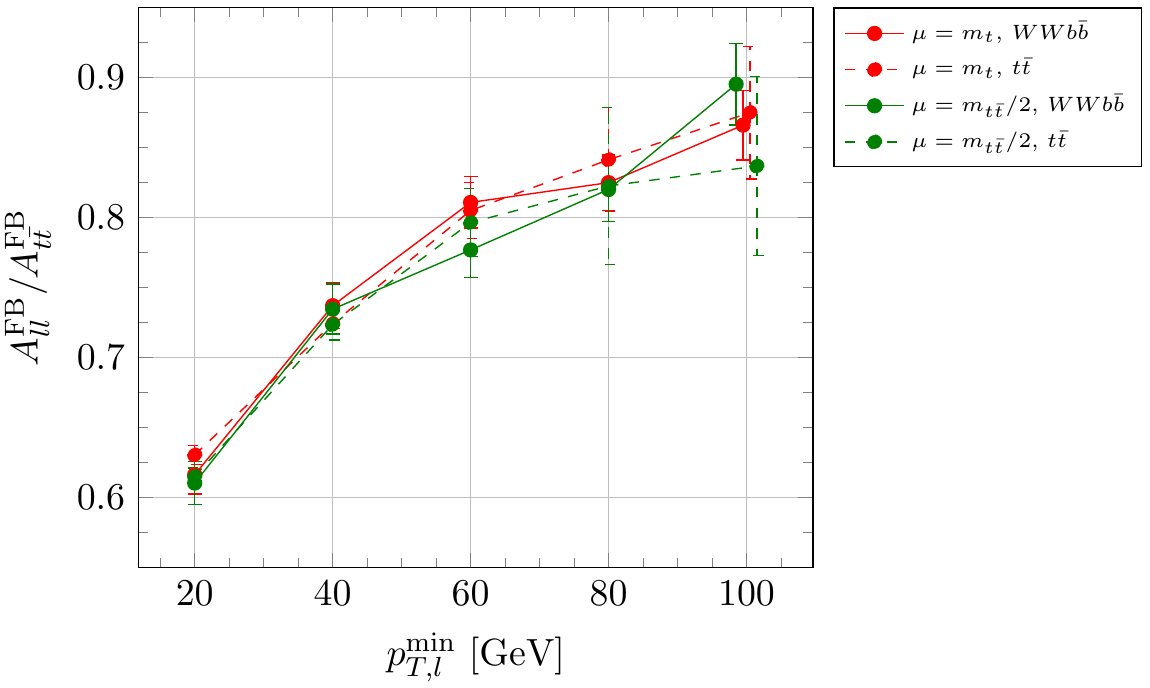}
  \end{subfigure}
  \caption{\label{fig:ratio}
    Ratio between the leptonic and top quark forward-backward
    asymmetries at the Tevatron, as a function of $p^\srm{min}_{T,l}$.
    The solid (dashed) lines represent the outcomes of the full
    (factorized) QCD NLO corrections to $W^+W^-b\bar b$ final states
    contributing to the dilepton channel at
    $\mathcal{O}(\alpha^2_\mrm{s}\alpha^2)$. The vertical bars denote the
    Monte Carlo statistical uncertainties. Note that the predictions
    associated with the highest $p^\srm{min}_{T,l}$ requirement suffer
    from low statistics. For better visibility of the individual
    results, these points therefore have been slightly shifted along
    the horizontal axis.}
\end{figure}

The ratio $A^\srm{FB}_{ll}/A^\srm{FB}_{t\bar t}$ is known to be less
affected by scale choices and uncertainties. We therefore show 
these ratios for all our different choices in
Figure~\ref{fig:ratio}. Indeed we find these ratio predictions to be
more robust, and conclude that this quantity can be predicted more
reliably than the absolute behaviour of the asymmetries. Again, the
trend of increasing correlation between the two types of asymmetries
can be seen for events containing more strongly boosted leptons.

%%%%%%%%%%%%%%%%%%%%%%%%%%%%%%%%%%%%%%%%%%%%%%%%%%

%\clearpage
\section{Conclusions}\label{sec:conclusion}

We have calculated the NLO QCD corrections to the processes
$pp\,(p\bar{p})\rightarrow W^+W^-b\bar{b}\to
(e^+ \nu_e)\,(\mu^- \bar{\nu}_{\mu})\,b\bar{b}$ in the 5-flavour
scheme, including non-resonant diagrams and singly resonant top quark
contributions, using the automated one-loop generator \tsc{GoSam} in
combination with the Monte Carlo program \tsc{Sherpa}.
We also performed an NLO calculation of top quark pair production in
the narrow width approximation supplemented by LO top quark decays,
enabling us to assess the impact of the non-factorizing contributions
at NLO. We found a reduction of the scale dependence of the total
cross section from about $30\%$ at LO to about $5\%$ at NLO,  
and a significant impact of the non-factorizing contributions on the
shape of the distribution for certain observables, for example the
invariant mass of a lepton and a $b$ jet, \mlb.

\smallskip
We also presented a detailed study of NLO effects in \mt\ measurements
based on the \mlb\ observable, making contact to a recent ATLAS
analysis, which uses a template method~\cite{ATLAS-CONF-2013-077}.
%
%We observed that the shape distortions in the \mlb\ distribution due
%to the NLO corrections for the full approach are substantial, while
%they are small for the factorized approach.
%
Using the factorized calculation, we observed only small shape
distortions in the \mlb\ distribution originating from NLO corrections.
The situation changes for the full approach where such distortions
turn out to be substantial.
The size of the resulting parton level prediction for the \mt offset,
when using LO templates as the theory model, was investigated using a
pseudo-data parton level analysis closely following the ATLAS
strategy.
For the full approach, the offset was found to be non-negligible,
amounting to about $1.9\gev$. This has to be contrasted with a
considerably smaller offset of about $0.5\gev$ in the factorized
approach.
In addition, we estimated the uncertainty on \mt\ resulting from
standard NLO scale variations. For the full approach, it was found to
be of the order of $1\gev$. This is much larger than the small
uncertainty of about $0.2\gev$, which we evaluated for the factorized
approach.
Clearly, further investigations using fully simulated events are
needed to properly assess the corresponding uncertainty on \mt\ within
the experimental analysis.
However, given our findings, and considering the fact that presently,
the experimental analyses are based on the factorized approach,
a larger uncertainty is likely to be found.

\smallskip
Finally, we focused on a study of top quark asymmetries where
we investigated the impact of the NLO corrections to $W^+W^-b\bar b$
production on both the top quark and leptonic charge asymmetry at the
LHC, and the top quark and leptonic forward-backward asymmetry at the
Tevatron. Our study centered on a more detailed investigation of the
correlation between the top quark and the leptonic forward-backward
asymmetries. In particular, we showed how this correlation changes as
a function of the kinematic requirement on the minimum transverse
momentum of the charged leptons, $p^\srm{min}_{T,l}$, quantifying its
sensitivity to different scale choices. The difference between the top
quark and leptonic asymmetry is observed to decrease for increasing
$p^\srm{min}_{T,l}$. While the individual absolute asymmetries depend
rather strongly on the employed scale choice, the ratio
$A^\srm{FB}_{ll}/A^\srm{FB}_{t\bar t}$ is found to be less sensitive
under such variations. Invoking the factorized approach, the NLO
corrections to $W^+W^-$ production in association with two $b$ jets
yield smaller asymmetry values throughout. This reduction is not seen
for the asymmetry ratios. The effect drops out and we obtain
predictions similar to those of the full approach.

%%%%%%%%%%%%%%%%%%%%%%%%%%%%%%%%%%%%%%%%%%%%%%%%%

\section*{Acknowledgements}

We would like to thank Stefan H\"oche for numerous discussions and
help with \tsc{Sherpa} generator issues.
We are grateful to Markus Schulze for helpful comparisons.
We also thank Gionata Luisoni, Nicolas Greiner, Stefano Pozzorini and
Jay Wacker for fruitful discussions. We are also grateful to the
\tsc{GoSam} collaboration for their work on code development.
%We further acknowledge the support of the Research Executive Agency (REA)
%of the European Union under the Grant Agreement number
%PITN-GA-2010-264564 (LHCPhenoNet).

%%%%%%%%%%%%%%%%%%%%%%%%%%%%%%%%%%%%%%%%%%%%%%%%%

%\appendix

%%%%%%%%%%%%%%%%%%%%%%%%%%%%%%%%%%%%%%%%%%%%%%%%%
%% REFS %%%
%%%%%%%%%%%%%%%%%%%%%%%%%%%%%%%%%%%%%%%%%%%%%%%%%

%\newpage
%\bibliographystyle{elsart-num} 
\bibliographystyle{JHEP}
\bibliography{refs_wwbb}

%%%%%%%%%%%%%%%%%%%%%%%%%%%%%%%%%%%%%%%%%%%%%%%%%
\end{document}